\pdfoutput=1
\documentclass[12pt]{article}
\usepackage{graphicx}
\begin{document}
\begin{titlepage}
\title{Tetrad Fields, Reference Frames, and the 
Gravitational Energy-Momentum in the 
Teleparallel Equivalent of General Relativity}

\author{J. W. Maluf$\,^{(a)(1)}$, F. L. Carneiro$\,^{(b)(2)}$, \\
\bigskip
S. C. Ulhoa$\,^{(c)(1)}$ and J. F. da Rocha-Neto$\,^{(d)(1)}$ \\
{(1)}\, Universidade de Bras\'{\i}lia, Instituto de F\'{\i}sica, \\
70.910-900 Bras\'{\i}lia DF, Brazil\\
(2)\, Universidade Federal do Norte do Tocantins, \\ 
Centro de Ci\^encias Integradas, 77824-838 Aragua\'ina
TO, Brazil  }
\date{}
\maketitle
\begin{abstract}
We review the concept and definitions of the energy-momentum and angular
momentum of the gravitational field in the teleparallel equivalent of
general relativity (TEGR). The importance of these definitions is justified
by three major reasons. First, the TEGR is a well established and widely
accepted formulation of the gravitational field, whose basic field strength
is the torsion tensor of the Weitzenb\"ock connection. Second, in the phase
space of the TEGR there exists an algebra of the Poincaré group. Not only
the definitions of the gravitational energy-momentum and 4-angular momentum
satisfy this algebra, but also the first class constraints related to these
definitions satisfy the algebra. And third, numerous applications of
these definitions lead to physically consistent results. These definitions
follow from a well established Hamiltonian formulation, and rely on the idea
of localization of the gravitational energy. In this review we revisit the
concept of localizability of the gravitational energy, in light of results
obtained in recent years. We have studied the behaviour of free particles in
the space-time of plane fronted gravitational waves (pp-waves). Free
particles are here understood as particles that are not subject to external
forces other than the gravitational acceleration due to pp-waves. Since
these particles acquire or loose kinetic energy locally, the transfer
of energy from or to the gravitational field must also be localized. 
We consider this theoretical result an important and definite argument in 
favour of the localization of the gravitational energy-momentum, and by 
extension, of the gravitational 4-angular momentum.
\end{abstract}
\thispagestyle{empty}
\vskip 1.0cm
\thispagestyle{empty}

\noindent (a) jwmaluf@gmail.com, wadih@unb.br\par
\noindent (b) fernandolessa45@gmail.com\par
\noindent (c) sc.ulhoa@gmail.com\par
\noindent (d) jfrocha@unb.br, jfrn74@yahoo.com \par

\vskip 1.0cm

\end{titlepage}
\tableofcontents

\section{Introduction}
The concept of energy is fundamental and indispensable 
in physics. Energy is a quantity that can {\it potentially} produce motion
of particles or physical bodies. We recall the statement by Galileo, 
according to which {\it ``My purpose is to set forth a very new science 
dealing with a very ancient subject. There is, in nature, perhaps nothing 
older than motion"}, which implies that energy is, likewise, one of the 
oldest manifestations in nature. Energy is an attribute of basically all 
particles and fields in the universe, but the energy of the gravitational 
field is, surprisingly, an issue whose 
concept and definition are still in debate and discussion in the physics 
community. There are various approaches to the definition of the energy
of the gravitational field, in part because there are several descriptions of
the relativistic gravitational field that are equivalent to each other, and 
also in part because there are misconceptions about the idea of localization
of the gravitational energy. All (or most) the descriptions mentioned above 
lead (or are equivalent) to Einstein's field equations,
and for this reason there is an ongoing debate about which definitions for
the energy, momentum and 4-angular momentum of the gravitational field are
correct.

The traditional and dominant view of the concept of gravitational energy 
takes place in the metric formulation of general relativity. According to 
this view, there does not exist an energy-momentum (or stress-energy) tensor
for the gravitational field, because the gravitational field ``can always
be locally set to zero", and therefore the gravitational energy cannot be
localized. The argument is the following.\par 
\bigskip

\noindent {\bf I.}  It is always possible to choose a ``locally inertial 
frame" (this is essentially a choice of coordinates) where the gravitational
``force" vanishes.\par 
\bigskip
\noindent {\bf II.}
This result is accomplished by arguing that the Christoffel symbols
vanish on a space-time event or over a timelike geodesic, 
and thus the metric tensor is locally flat. \par 
\bigskip
\noindent {\bf III.}
Finally, it is argued that this vanishing is a
manifestation of the principle of equivalence in general relativity. Note
that, here, the manifestation of the principle of equivalence is due to 
(and obtained by) a choice of coordinates, despite the fact that a choice of
coordinates on an arbitrary manifold should not have relevant physical 
(dynamical) consequences.\par  
\bigskip

As a result, many physicists adhere (or feel comfortable) to the use of 
pseudo-tensors, of which there are numerous definitions in the literature.
Pseudo-tensors are used in many analyses of the energy of gravitational 
waves. The difficulties mentioned above regarding a definition for the
stress-energy tensor of the gravitational field 
gave rise to the concept of quasi-local {\it mass}, which is related 
to the energy enclosed by a 2-surface in space. This quantity, in turn,
is not associated to any form of energy density, or scalar density on the
manifold. The applicability of the
definition of quasi-local gravitational mass has its own difficulties, as
we will discuss ahead in this review. It must be emphasized that the 
vanishing mentioned above of the Christoffel symbols takes place on a
space-time event or, more generally, along any geodesic, timelike or 
spacelike geodesic. 

The fact that we may set the Christoffel symbols to vanish on a 
spacelike geodesic means that this vanishing is a just feature of
differential geometry \cite{Grishchuk}.
It is certainly not a consequence of any principle 
in physics, including the principle of equivalence. In the 
teleparallel equivalent of general relativity (TEGR), the Christoffel
symbols may be constructed, of course,
but they do not play any major role in the theory. They
may be defined as a secondary, or auxiliary concept. The field strength
of the TEGR is the torsion tensor of the Weitzenb\"ock connection, and the 
torsion tensor cannot be made to vanish by means of a coordinate 
transformation. 

Einstein's formulation of the principle of equivalence has 
been discussed at length by Norton \cite{Norton}, and readdressed in Ref.
\cite{Maluf0}. In summary, we just highlight two difficulties presented in
Ref. \cite{Norton}, that allows to
contradict the statements {\bf I}, {\bf II}, {\bf III} above regarding the
impossibility of having a localized definition for the gravitational energy.
The first difficulty is that the version of the principle
of equivalence that is widely adopted in the literature, as the one displayed
above, is due to 
Pauli \cite{Norton}, and is rather distinct from Einstein's version of the
principle. According to Pauli’s formulation, ``in every infinitely small 
neighborhood of the space-time, there always exists a coordinate system in
which gravitation has no influence either on the motion of particles or any 
other physical processes". This conclusion is justified by the vanishing of
Christoffel symbols at a point within this neighborhood. But we know that 
what really vanishes at this point are the first derivatives of the metric
tensor, not the second or higher derivatives. Thus, the vanishing of the
gravitational field even at the space-time point considered is questionable.
We quote here a statement from Ref. \cite{Norton} (and reproduced in 
Ref. \cite{Maluf0}):\par 
\bigskip

{\it It has rarely been acknowledged
that Einstein never endorsed the principle of equivalence which results,
here called the ``infinitesimal principle of equivalence". Moreover, his 
early correspondence} (with Pauli) {\it contains a devastating objection to 
this principle: in infinitesimal regions of the space-time manifold it is 
impossible to distinguish geodesics from many other curves and therefore 
impossible to decide whether a point mass is in free fall}. \par 
\bigskip
\noindent Thus, in infinitesimal regions of the space-time, it is impossible
to assert whether gravitation has or has not influence on the motion of free
particles.

The second difficulty presented in Ref. \cite{Norton} is that Einstein 
considered the principle of equivalence as a consequence of frame 
transformations, not coordinate transformations. His formulation may be 
summarized as follows  \cite{Norton}.\par 
\bigskip

{\it Consider a reference frame K (a Galilean system) and a reference frame 
K', which is uniformly accelerated with respect to K. Then one asks whether
an observer in K' must understand his condition as accelerated, or whether 
there remains a point of view according to which he can interpret his 
condition as at ``rest". Einstein concludes that by assuming the existence of
a homogeneous gravitational field in K' it is possible to consider the latter 
as at rest. In his words: The assumption that one may treat K' as at rest, in 
all strictness without any laws of nature not being fulfilled with respect 
K', I call the Principle of Equivalence.}\par 
\bigskip

The concepts of frame transformations and coordinate transformations are very
much clear in general relativity, and are different from each other. Frame
transformations act on tetrad fields, for instance. {\it ``The happiest 
thought of my life"}, asserted by Einstein and thoroughly described in the 
book by Schucking and Surowitz \cite{Schucking}, is realized by means of a 
frame transformation, it is an actual free fall, and definitely is 
not a coordinate transformation. The principle of 
equivalence is a very beautiful idea, but the eminent physicist Synge
\cite{Synge} claimed that he does not understand it as a principle for
general relativity, because either there is or there is not a gravitational
field in any arbitrary neighborhood of the space-time, since other relevant
tensors such as the Riemann-Christoffel or torsion tensors do not vanish as a
consequence of a coordinate transformation. Synge's argument is logically
sound. The existence or removal of the gravitational field cannot depend on
a choice of coordinates.

In conclusion, the arguments {\bf I}, {\bf II} and {\bf III}
presented above regarding the impossibility of the existence of a
localized expression for the gravitational energy are at the same time very 
weak and insufficient, and place no impediment for the localization of the
gravitational energy-momentum.

This review addresses the gravitational energy-momentum and 4-angular 
momentum in the context of the TEGR. This theory has been  
investigated by several authors over the years, see Refs. 
\cite{Hayashi}-\cite{Formiga9}. It turns out that the TEGR is a suitable
formulation for the relativistic gravitational field, that yields 
interesting results in cosmology, in the framework of modified $f(T)$ 
theories \cite{Ferraro1}-\cite{Capozziello1}. In this review we
will adopt the definitions and notations of Ref. \cite {Maluf3}. In recent 
times, we have presented several results in the literature regarding
the consistency and plausibility of the definitions of energy-momentum and
4-angular momentum of the gravitational field in the TEGR. All definitions 
are based on densities which are well behaved under coordinate 
transformations, and therefore our definitions are localized. 

The definitions in consideration here do not rely on abstract mathematical
elaborations whose purpose is to yield the Schwarzschild mass, for instance.
Rather, they follow
from the field equations of the TEGR. The Lagrangian field equations 
allow the establishment of the gravitational energy-momentum
4-vector $P^a$, the flux (balance) equations and the fluxes definitions,
but the complete structure formed by the gravitational
energy-momentum and 4-angular momentum is obtained in the realm of the
Hamiltonian formulation. As we will discuss ahead, the 6 extra degrees of
freedom of the tetrad field (which has 16 a priori degrees of freedom,
compared to the 10 degrees of freedom of the metric tensor) allow the
definition of the 3-angular momentum and of the 3-centre of mass momentum of 
the gravitational field. These latter definitions follow from the Hamiltonian
first class constraints of the theory. Thus, all definitions to be considered
in this review are consequence of the Hamiltonian field equations of the 
theory, as we will show.

In recent years, we have investigated the trajectories of idealised free
particles in the presence of plane-fronted gravitational waves (pp-waves).
The pp-waves are very simple solutions of Einstein's field equations in 
vacuum, even simpler than the Schwarzschild solution. The idealised particles
in consideration are particles with negligible mass, that do not
influence the gravitational field around them, and are ``free" in
the sense that they are subject only to the gravitational field of the 
pp-waves, i.e., they follow geodesics in space-time.
In a series of articles \cite{Maluf4,Maluf5,Maluf6,Maluf7} we found
that the free particles undergo a velocity memory effect, and also that  
they gain or loose kinetic energy after the passage of the pp-wave. Since
the only physical entity in question is the pp-wave, there is an energy
transfer between the particles and the gravitational field of the waves.
Evidently, this energy transfer takes place at the position of the particle,
and thus the gravitational energy transfer also takes place at the 
localization of the pointwise particle. It is obvious, therefore, that energy
of the gravitational field of the pp-wave must be localized. It does not make
sense to relate this energy transfer to closed spatial 2-surfaces of 
arbitrary radius around the particle. 
Needless to say, it is even more absurd to relate the
energy transfer to the information storaged on a 2-surface at spatial 
infinity, if the total gravitational energy in question were an ADM type 
energy \cite{ADM}, for instance.

We consider this feature of the physical system pp-wave + idealised particle
as a strong and definite argument in favour of the localization of the
gravitational energy. This argument holds true also in the context of 
linearised gravitational waves, where the velocity memory effect is also
predicted to take place, but which is not considered here.

The plan of this review is the following. In Section 2 we will review the 
concept of tetrad fields as frames adapted to arbitrary observers in 
space-time. This interpretation is the primary attribute of tetrad fields. 
Contrary to many applications in the literature, where tetrad fields are
treated as just a collection of four 
orthonormal 4-vectors in space-time, here the tetrad
fields are adapted (attached) to the worldline of an observer. The inevitable
conclusion is that tetrad fields provide information of both the observer
and the gravitational field. In Section 3
we present both the Lagrangian and Hamiltonian formulations of the TEGR, the
definitions of the energy-momentum and 4-angular momentum of the 
gravitational field, and the Poincar\'e algebra between these definitions,
in the phase space of the theory.
To our knowledge, the Hamiltonian formulation of the TEGR is the only
framework where the algebra of the Poincar\'e group is realized by means of 
the Poisson brackets between the energy-momentum and 4-angular momentum 
definitions.

We will review only three major applications of the definitions of 
gravitational energy-momentum. In Section 4 we will readdress the 
gravitational energy enclosed by the external event horizon of the Kerr black
hole. This energy has also been investigated by the quasi-local approach. 
This is
an opportunity to compare our result with the latter and to show that the
result obtained in the TEGR is quite remarkable. We will show that energy
contained within the external event horizon of the Kerr black hole is 
extremely close to $2M_{irr}$, where $M_{irr}$ is the irreducible mass of the
black hole.

In Section 5 we will consider the Schwarzschild-de Sitter space-time, with 
the purpose of analysing the distribution of gravitational energy in the 
vicinity of the cosmological constant $\Lambda$. The sharp increase of the 
gravitational energy density as one approaches the cosmological horizon is
a feature that shares similarities with the dark energy 
problem, in spite of the fact that we are considering a static physical
configuration. In Section
6 we readdress the Bondi-Sachs space-time, characterised by the mass aspect 
$M$ and by two functions, $c$ and $d$. The two news functions 
$\partial_0 c$ and $\partial_0 d$ carry 
information both at spacelike and null infinities, and therefore they should 
contribute to the total energy of the space-time. That is indeed what we 
find in the context of the TEGR. The total energy of the Bondi-Sachs 
space-time is normally given by the mass 
function $m(u)$ only, which is an integral of the mass aspect, 
but here we show that the news functions also 
contribute, since they are manifestations of the Bondi-Sachs line element
at spatial infinity. Section 7 is devoted to the study of geodesics of free
particles in the space-time of plane-fronted gravitational waves. Only few
examples suffice to demonstrate that the kinetic energy of free particles 
increases or decreases with the passage of some simple constructions of 
pp-waves. The results of this section not only support the idea of 
localization of gravitational energy, but they show that the latter is 
mandatory. And finally, in Section 8 we present our conclusions. \par 
\bigskip
\noindent Notation:

\begin{enumerate}
\item
Space-time indices $\mu, \nu, ...$ and SO(3,1) (Lorentz) indices
$a, b, ...$ run from 0 to 3. Time and space indices are indicated according to
$\mu=0,i,\;\;a=(0),(i)$.

\item
The tetrad fields are represented by $e^a\,_\mu$, and the torsion tensor by
$T_{a\mu\nu}=\partial_\mu e_{a\nu}-\partial_\nu e_{a\mu}$.
The flat, tangent space Minkowski 
space-time metric tensor raises and lowers tetrad indices, and is fixed
by $\eta_{ab}= e_{a\mu} e_{b\nu}g^{\mu\nu}=(-1,+1,+1,+1)$.
\item
The frame components are given by the inverse tetrads 
$\lbrace e_a\,^\mu \rbrace$,
although we may as well refer to $\lbrace e^a\,_\mu\rbrace$ as the frame. The
determinant of the tetrad fields is represented by $e=\det(e^a\,_\mu)$
\end{enumerate}

The torsion tensor defined above is often related to the object of
anholonomity $\Omega^\lambda\,_{\mu\nu}$ via 
$\Omega^\lambda\,_{\mu\nu}= e_a\,^\lambda T^a\,_{\mu\nu}$.
However, we assume that the space-time geometry is defined by the 
tetrad fields only, and in this case the only
possible nontrivial definition for the torsion tensor is given by
$T^a\,_{\mu\nu}$. This torsion tensor is related to the 
antisymmetric part of the Weitzenb\"ock  connection 
$\Gamma^\lambda_{\mu\nu}=e^{a\lambda}\partial_\mu e_{a\nu}$, which
establishes the Weitzenb\"ock space-time. 
The metric and torsion-free Christoffel symbols are denoted by 
$^0\Gamma^\lambda_{\mu\nu}$, and the associated torsion-free Levi-Civita
connection by $^0\omega_{\mu ab}$, in the equations below. The parameter
$c$ denotes the speed of light, but sometimes it will be omitted by
assuming $c=1$.

\section{Tetrad fields as reference frames in space-time}

A set of tetrad fields is a collection of four orthonormal, linearly 
independent vector fields in space-time,
$\lbrace e^{(0)}\,_\mu, e^{(1)}\,_\mu, e^{(2)}\,_\mu, e^{(3)}\,_\mu\rbrace$.
They may constitute the local reference frame of an observer that moves along
a trajectory $C$, represented by the worldline $x^\mu(\tau)$ 
\cite{Hehl3,Maluf8,Maluf9}. The parameter $\tau$ is the proper time of the 
observer. 
The components $e^{(0)}\,_\mu$ and $e^{(i)}\,_\mu$ are timelike and spacelike
vectors, respectively; $e^a\,_\mu$ transforms as covariant vector fields under
coordinate transformations of the space-time manifold, and as contravariant 
vector fields under SO(3,1) (Lorentz) transformations, i.e., 
$\tilde{e}^a\,_\mu=\Lambda^a\,_b\,e^b\,_\mu$,
where the matrices $\lbrace \Lambda^a\,_b\rbrace$ are representations of the
SO(3,1) group and satisfy $\Lambda^a\,_c\Lambda^b\,_d\, \eta_{ab}=\eta_{cd}$.
The metric tensor $g_{\mu\nu}$ is obtained by the standard 
relation $e^a\,_\mu e^b\,_\nu \eta_{ab}=g_{\mu\nu}$.

Tetrad fields $e^a\,_\mu$ allow the projection of vectors and tensors in 
space-time in the local frame of an observer. Vectors and tensors components
are abstract, space-time dependent quantities.
In order to measure field quantities with magnitude and direction, an observer
must project these quantities on the frame 
carried by the observer. The projection of a vector $V^{\mu}(x)$ at a position
$x^\mu$, on a
particular frame, is simply given by by 
$V^{a}(x) = e^{a}\,_{\mu}(x)\,V^{\mu}(x)$.

Given a  worldline $C$ of an observer, represented by 
$x^\mu(\tau)$, the velocity of the observer along $C$ is denoted by 
$u^\mu(\tau)=dx^\mu/d\tau$. We identify the observer's
velocity with the $a=(0)$ component of $e_a\,^\mu$. Thus,
$e_{(0)}\,^\mu=u^\mu(\tau)/c$. The acceleration $a^\mu$ of the observer
is given by the absolute derivative of $u^\mu$ along $C$, 

\begin{equation}
a^\mu= {{Du^\mu}\over{d\tau}} =c{{De_{(0)}\,^\mu}\over {d\tau}} =
c\,u^\alpha \nabla_\alpha e_{(0)}\,^\mu\,, 
\label{1-1}
\end{equation}
where the covariant derivative is constructed out of the Christoffel symbols
$^0\Gamma^\mu_{\alpha\beta}$. The last equality is obtained as follows,

\begin{eqnarray}
{{De_{(0)}\,^\mu}\over {d\tau}}&=& {{de_{(0)}\,^\mu}\over {d\tau}}+\,\,
^0\Gamma^\mu_{\alpha\beta}\, {{dx^\alpha}\over{d \tau}}\, e_{(0)}\,^\beta
\nonumber \\
&=& {{dx^\alpha} \over {d\tau}}\,
{{\partial e_{(0)}\,^\mu}\over {\partial x^\alpha}} +\,\,
^0\Gamma^\mu_{\alpha\beta}\, {{dx^\alpha}\over{d \tau}}\, e_{(0)}\,^\beta
\nonumber  \\
&=& u^\alpha \,\nabla _\alpha e_{(0)}\,^\mu\,.
\label{1-2}
\end{eqnarray}

Thus, $e_a\,^\mu$ allows to obtain 
the velocity and acceleration of an observer along 
the worldline. Therefore, for a given set of tetrad fields, 
$e_{(0)}\,^\mu$ describes a congruence of timelike curves, which determines
a field of observers. These observers are characterized by the velocity field
$u^\mu=c\,e_{(0)}\,^\mu$, and are endowed with acceleration $a^\mu$. 
If $e^a\,_\mu \rightarrow \delta^a_\mu$ in the spacelike limit 
$r \rightarrow \infty$, then $e^a\,_\mu$ is adapted to static
observers at spatial infinity. 

The geometrical characterization of the whole set of 
tetrad fields as an observer's frame 
may be given by considering the acceleration of the frame along an 
arbitrary path $x^\mu(\tau)$ of an observer. The acceleration 
of the whole frame is determined by the absolute derivative of $e_a\,^\mu$ 
along $x^\mu(\tau)$. Thus, assuming that the observer carries an orthonormal 
tetrad  frame $e_a\,^\mu$, the acceleration of the frame along the path is 
given by \cite{Mashh2,Mashh3}

\begin{equation}
{{D e_a\,^\mu} \over {d\tau}}=\phi_a\,^b\,e_b\,^\mu\,,
\label{1-3}
\end{equation}
where $\phi_{ab}$ is the antisymmetric acceleration tensor. According to Refs. 
\cite{Mashh2,Mashh3}, in analogy with the Faraday tensor we may identify 
$\phi_{ab} \rightarrow ({\bf a}/c, {\bf \Omega})$, where ${\bf a}$ is the 
translational acceleration ($\phi_{(0)(i)}=a_{(i)}/c$) and ${\bf \Omega}$ is the 
frequency of rotation of the local spatial frame with respect to a 
non-rotating, Fermi-Walker transported frame. 
It follows from Eq. (\ref{1-3}) that

\begin{equation}
\phi_a\,^b= e^b\,_\mu {{D e_a\,^\mu} \over {d\tau}}=
e^b\,_\mu \,u^\lambda\nabla_\lambda e_a\,^\mu\,.
\label{1-4}
\end{equation}

The acceleration vector $a^\mu$ defined by Eq. (\ref{1-1}) may be projected
on a frame. The result is

\begin{equation}
a^b= e^b\,_\mu a^\mu=c\,e^b\,_\mu u^\alpha \nabla_\alpha
e_{(0)}\,^\mu=c\,\phi_{(0)}\,^b\,.
\label{1-5}
\end{equation}
Thus, $a^\mu$ and $\phi_{(0)(i)}$ are not different translational 
accelerations of the frame. 

The expression of $a^\mu$ given by Eq. (\ref{1-1}) may be rewritten as

\begin{eqnarray}
a^\mu/c&=& u^\alpha \nabla_\alpha e_{(0)}\,^\mu 
=u^\alpha \nabla_\alpha u^\mu =
{{dx^\alpha}\over {d\tau}}\biggl(
{{\partial u^\mu}\over{\partial x^\alpha}}
+\,\,^0\Gamma^\mu_{\alpha\beta}u^\beta \biggr) \nonumber \\
&=&{{d^2 x^\mu}\over {d\tau^2}}+\,\,^0\Gamma^\mu_{\alpha\beta}
{{dx^\alpha}\over{d\tau}} {{dx^\beta}\over{d\tau}}\,,
\label{1-6}
\end{eqnarray}
where $\,\,^0\Gamma^\mu_{\alpha\beta}$ are the Christoffel symbols.
We see that if $u^\mu=c\,e_{(0)}\,^\mu$ represents a geodesic
trajectory, then the frame is in free fall and 
$a^\mu/c=0=\phi_{(0)(i)}$. Therefore we conclude that non-vanishing
values of the latter quantities do represent {\it inertial} 
(i.e., non-gravitational) accelerations of the frame. A static frame in an
asymptotically flat space-time is an example of a frame that is subject to
inertial accelerations.

Since the tetrads are orthonormal vectors, we may  rewrite Eq. 
(\ref{1-4}) as
$\phi_a\,^b= -u^\lambda e_a\,^\mu \nabla_\lambda e^b\,_\mu$, 
where $\nabla_\lambda e^b\,_\mu=\partial_\lambda e^b\,_\mu-\,\,
^0\Gamma^\sigma_{\lambda \mu} e^b\,_\sigma$. Now we take into account
the identity 

$$\partial_\lambda e^b\,_\mu-\,\,
^0\Gamma^\sigma_{\lambda \mu} e^b\,_\sigma+\,\,
^0\omega_\lambda\,^b\,_c e^c\,_\mu \equiv 0\,,$$ 
where $^0\omega_\lambda\,^b\,_c$ is the metric compatible, torsion-free
Levi-Civita connection, 

\begin{eqnarray}
^0\omega_{\mu ab}&=&-{1\over 2}e^c\,_\mu(
\Omega_{abc}-\Omega_{bac}-\Omega_{cab})\,,  \nonumber \\
\Omega_{abc}&=&e_{a\nu}(e_b\,^\mu\partial_\mu
e_c\,^\nu-e_c\,^\mu\partial_\mu e_b\,^\nu)\,,
\label{1-7}
\end{eqnarray}
and express $\phi_a\,^b$ as

\begin{equation}
\phi_a\,^b=c\,e_{(0)}\,^\mu(\,\,^0\omega_\mu\,^b\,_a)\,.
\label{1-8}
\end{equation}
At last we consider the {\it identity} $\,\,^0\omega_\mu\,^a\,_b=
-K_\mu\,^a\,_b$, where $K_\mu\,^a\,_b$ is the 
contortion tensor defined by

\begin{equation}
K_{\mu ab}={1\over 2}e_a\,^\lambda e_b\,^\nu(T_{\lambda \mu\nu}+
T_{\nu\lambda\mu}+T_{\mu\lambda\nu})\,,
\label{1-9}
\end{equation}
and $T_{\lambda \mu\nu}=e^a\,_\lambda T_{a\mu\nu}$ (see  
Eq. (4) of Ref. \cite{Maluf10}; the identity may be obtained by direct
calculation). After simple manipulations, we finally obtain

\begin{equation}
\phi_{ab}={c\over 2} \lbrack T_{(0)ab}+T_{a(0)b}-T_{b(0)a}
\rbrack =-\phi_{ba}\,.
\label{1-10}
\end{equation}
The expression above is clearly not invariant under local SO(3,1)
transformations, but is invariant under coordinate transformations.

The values of $\phi_{ab}$ for given tetrad fields 
may be used to characterize the frame. We interpret $\phi_{ab}$ as the
inertial accelerations along the trajectory $x^\mu(\tau)$.
Therefore, given any set of tetrad fields for an arbitrary space-time, its 
geometrical interpretation may be obtained,\par 
\bigskip
\noindent  {\bf (i)} either by suitably 
identifying $e_{(0)}\,^{\mu}=u^\mu/c$, 
together with the orientation in the three-dimensional space of 
the components $e_{(1)}\,^\mu,e_{(2)}\,^\mu,e_{(3)}\,^\mu$, considering the
symmetry of the physical configuration, or,\par 
\bigskip
\noindent  {\bf (ii)} by the values of the 
acceleration tensor $\phi_{ab}=-\phi_{ba}$, which characterize
the inertial state of the frame.\par 
\bigskip
\noindent  The condition $e_{(0)}\,^{\mu}=u^\mu/c$ fixes
only the three components $e_{(0)}\,^1$, $e_{(0)}\,^2$, $e_{(0)}\,^3$, because
the component $e_{(0)}\,^0$ is determined by the normalization condition 
$u^\mu u^\nu g_{\mu\nu}=-c^2$. In both cases, the fixation of the frame 
requires the fixation of 6 components of the tetrad fields.\par 
\bigskip

Two simple, familiar and straightforward applications of Eqs. (1-10) are the
following:\par
\bigskip

\noindent {\bf (1)}
The tetrad fields adapted to observers at rest in Minkowski space-time is given
by $e^a\,_\mu(ct,x,y,z)=\delta^a _\mu$. By means of a time-dependent boost in 
the $x$ direction, say, the tetrad fields read

\begin{equation}
e^a\,_\mu(ct,x,y,z)=\pmatrix{\gamma&-\beta\gamma&0&0\cr
-\beta\gamma&\gamma&0&0\cr
0&0&1&0\cr
0&0&0&1\cr}\,,
\label{1-11}
\end{equation}
where $\gamma=(1-\beta^2)^{-1/2}$, $\beta=v/c$ and $v=v(t)$. In the expression
above $a$ and $\mu$ label rows and columns, respectively. The
frame above is adapted to observers whose four-velocity is 
$u^\mu(t,x,y,z)=c\, e_{(0)}\,^\mu=(c\,\gamma, c\,\beta\gamma,0,0)$. After
simple calculations we obtain

\begin{eqnarray}
\phi_{(0)(1)}&=&c\,{d\over {dx^0}}\lbrack \beta \gamma\rbrack = 
{d \over {dt}}\biggl[
{ {v/c} \over {\sqrt{1-v^2/c^2} }} \biggr]\,, \\ \nonumber
\phi_{(0)(2)}&=&0 \,, \\ \nonumber
\phi_{(0)(3)}&=&0 \,,
\label{1-12}
\end{eqnarray}
and $\phi_{(i)(j)}=0$.\par 
\bigskip

\noindent {\bf (2)} The frame adapted to an observer at rest in Minkowski
space-time whose four-velocity is $u^\mu=(c,0,0,0)$, endowed with spatial 
axes that rotate around the $z$ axis, say, with angular velocity $\omega(t)$,
reads

\begin{equation}
e^a\,_\mu(ct,x,y,z)=\pmatrix{1&0&0&0\cr
0& \cos\omega(t)& -\sin\omega(t)&0\cr
0&\sin\omega(t)& \cos\omega(t)&0\cr
0&0&0&1\cr}\,.
\label{1-13}
\end{equation}
After simple calculations we obtain

\begin{eqnarray}
\phi_{(2)(3)}&=&0\,, \\ \nonumber
\phi_{(3)(1)}&=&0 \,, \\ \nonumber
\phi_{(1)(2)}&=&-c\, {{d\omega} \over {dx^0}} \,,
\label{1-14}
\end{eqnarray}
and $\phi_{(0)(i)}=0$. The examples above support the interpretation of 
$\phi_{ab}$ as the inertial accelerations of the frame.

\subsection{Fermi-Walker transported frames}

Fermi-Walker transported frames define a standard of non-rotation for 
accelerated observers. These are frames for which the frequency of rotation
$\phi_{(i)(j)}$ vanishes. The vanishing of $\phi_{(i)(j)}$ is the maximum
requirement one can impose on an arbitrary frame, in any space-time, 
regarding its rotational state.
Here we show how to transform an arbitrary frame 
into a Fermi-Walker transported frame, by means of a local Lorentz 
transformation. We will adopt the notation, presentation and results of Ref.
\cite{Synge}.

The absolute derivative of a vector $V^\mu$ along a world-line $C$ is defined
as usual by

\begin{equation}
{{D V^\mu}\over {d\tau}}={{dV^\mu}\over {d\tau}}+\,\,
^0\Gamma^\mu_{\alpha\beta} V^\alpha{{dx^\beta}\over{d\tau}}\,.
\label{1-2-1}
\end{equation}
We consider four 4-vectors, $A^\mu$, $B^\mu$, $C^\mu$ and $D^\mu$ that 
satisfy the following equations and properties:

\begin{eqnarray}
{{DA^\mu}\over {d\tau}}&=& b B^\mu 
\label{1-2-2} \,, \\
{{DB^\mu}\over {d\tau}}&=&cC^\mu+bA^\mu 
\label{1-2-3}  \,, \\
{{DC^\mu}\over {d\tau}}&=&dD^\mu-cB^\mu 
\label{1-2-4} \,, \\
{{DD^\mu}\over {d\tau}}&=&-dC^\mu \,, 
\label{1-2-5}
\end{eqnarray}
where $A^\mu A_\mu=-1$, $B^\mu B_\mu=C^\mu C_\mu=D^\mu D_\mu=1$, 
and $b,c,d$ are non-negative coefficients. Starting with the vector $A^\mu$,
Eq. (\ref{1-2-2}) defines $B^\mu$, Eq. (\ref{1-2-3}) defines $C^\mu$ and Eq. 
(\ref{1-2-4}) defines $D^\mu$. Equation (\ref{1-2-5}) is verified in view 
(i.e., it is a consequence) of 
Eqs. (\ref{1-2-2}-\ref{1-2-4}). It is not difficult to verify that Eqs. 
(\ref{1-2-2}-\ref{1-2-5}) imply that $A^\mu, B^\mu, C^\mu$ and $D^\mu$ 
form an orthonormal set of vectors, i.e., $A^\mu B_\mu =0$, etc.

We may identify $A^\mu$ with the unit vector tangent to the trajectory $C$,
$A^\mu=dx^\mu/ds$. In this case, $A^\mu, B^\mu, C^\mu$ and $D^\mu$ 
establish the Frenet-Serret frame, and Eqs. (\ref{1-2-2}-\ref{1-2-5}) 
are called the Frenet-Serret equations. $B^\mu$, $C^\mu$ and $D^\mu$ are the
first, second and third normals to $C$, and $b, c, d$ are the first, second
and third curvatures of $C$, respectively \cite{Synge}. Particular values of
the coefficients $b, c, d$ yield special curves in space-time. For instance,
if $b=c=d=0$, the curve $C$ is a geodesic; if $b=$ constant and $c=d=0$, $C$
represents a hyperbola, and if $b=$ constant, $c= $ constant and $d=0$, then
$C$ is a helix \cite{Synge}.

Let us consider a timelike trajectory $C$ represented by $x^\mu=x^\mu(\tau)$,
in a space-time determined by the metric tensor $g_{\mu\nu}$. The Fermi-Walker
transport of a vector $F^\mu$ along $C$ is defined by \cite{Synge}

\begin{equation}
{{D F^\mu}\over {d\tau}}=bF_\alpha (A^\mu B^\alpha -A^\alpha B^\mu)\,.
\label{1-2-6}
\end{equation}
Given the value $F^\mu(\tau_0)$ at a certain initial position $\tau_0$, Eq. 
(\ref{1-2-6}) allows the determination of $F^\mu$ along $C$. The Fermi-Walker
transport of a second rank tensor along $C$ is defined by

\begin{equation}
{{D T^{\mu\nu}}\over {d\tau}}=
bT_\alpha\,^\nu (A^\mu B^\alpha -A^\alpha B^\mu) +
bT^\mu\,_\alpha (A^\nu B^\alpha -A^\alpha B^\nu)\,.
\label{1-2-7}
\end{equation}
It follows from the equation above that 

\begin{equation}
{{D g^{\mu\nu}}\over {d\tau}}= 0={{D \delta^\mu_\nu}\over {d\tau}}\,.
\label{1-2-8}
\end{equation}
The velocity vector $A^\mu=dx^\mu/d\tau=u^\mu$ naturally undergoes 
Fermi-Walker transport. Application of Eq. (\ref{1-2-6}) to $A^\mu$ yields 
Eq. (\ref{1-2-2}). 

It is easy to see that the scalar product of two vectors is preserved
under the Fermi-Walker transport. Let $\phi$ represent the scalar
product of the vectors $\Sigma^\mu$ and $\Psi_\mu$. Along $C$ we 
have

\begin{eqnarray}
\phi(\tau + d\tau)-\phi(\tau)&=&\Sigma^\mu(\tau+d\tau)\Psi_\mu(\tau+d\tau)-
\Sigma^\mu(\tau)\Psi_\mu(\tau) \nonumber \\
{}&=& \Psi_\mu ( \delta^{FW}\Sigma^\mu) + 
\Sigma^\mu (\delta^{FW} \Psi_\mu)\,,
\label{1-2-9}
\end{eqnarray}
where

\begin{eqnarray}
\delta^{FW} \Sigma^\mu&=
&-\,\,^0\Gamma^\mu_{\alpha\beta}\Sigma^\beta dx^\alpha+
b\Sigma_\alpha(A^\mu B^\alpha-A^\alpha B^\mu)d\tau \,, \nonumber \\
\delta^{FW} \Psi_\mu&=&\,\,^0\Gamma^\lambda_{\alpha \mu} 
\Psi_\lambda dx^\alpha +
b\Psi_\alpha (A_\mu B^\alpha- A^\alpha B_\mu)d\tau \,.
\label{1-2-10}
\end{eqnarray}
Equations (\ref{1-2-9}) and (\ref{1-2-10}) lead to 
$\phi(\tau+d\tau)-\phi(\tau)=0$.

In what follows, we
identify  the velocity vector $A^\mu$ on $C$ with the timelike 
component of the tetrad fields $e_{(0)}\,^\mu$ according to

\begin{equation}
{1\over c}A^\mu= {1\over c}{{dx^\mu}\over {d\tau}}=e_{(0)}\,^\mu  \,.
\label{1-2-11}
\end{equation}
The important feature of this identification is that the
Fermi-Walker transport of $e_{(0)}\,^\mu$ along $C$
guarantees that $e_{(0)}\,^\mu$ will always be tangent to $C$. The
spacelike components $e_{(k)}\,^\mu$ are everywhere orthogonal to
$e_{(0)}\,^\mu$. Along $C$, the components $e_{(k)}\,^\mu$ also undergo
Fermi-Walker transport. Given that $e_{(k)}\,^\mu$ and $e_{(0)}\,^\mu$ 
are orthogonal, we have from Eq. (\ref{1-2-6}) that

\begin{equation}
{{D e_{(k)}\,^\mu}\over {d\tau}}=b e_{(0)}\,^\mu e_{(k)}\,^\lambda
B_\lambda\,.
\label{1-2-12}
\end{equation}
Taking into account Eq. (\ref{1-2-2}), the equation above may be rewritten as

\begin{equation}
{{D e_{(k)}\,^\mu}\over {d\tau}}=c\, e_{(0)}\,^\mu e_{(k)\lambda}
{{De_{(0)}\,^\lambda}\over {d\tau}}\,.
\label{1-2-13}
\end{equation}
In the equation above, $c$ is the speed of light. Equation (\ref{1-2-13})
determines the transport of the orthonormal basis $e_a\,^\mu$ along an 
arbitrary timelike curve $C$, such that $e_{(0)}\,^\mu$ is always tangent
to $C$.

In view of the definition given by Eq. (\ref{1-1}), Eq. (\ref{1-2-13}) may
be expressed as 

\begin{equation}
{{D e_{(k)}\,^\mu}\over {d\tau}}= u^\mu e_{(k)\lambda}a^\lambda \,.
\label{1-2-14}
\end{equation}
Making $b=(k)$ in Eq. (\ref{1-5}), we find 
$e_{(k)\lambda}a^\lambda=a_{(k)}=\phi_{(0)(k)}$. Thus the Fermi-Walker
transport of the frame may be written as

\begin{equation}
{{D e_{(k)}\,^\mu}\over {d\tau}}= u^\mu \phi_{(0)(k)}\,.
\label{1-2-15}
\end{equation}

On the other hand, it is easy to verify that the total acceleration
of the frame components $e_{(k)}\,^\mu$ given by Eq. (\ref{1-3}) may be
expressed in terms of $\phi_{(0)(k)}$ and $\phi_{(j)(k)}$ according to

\begin{equation}
{{D e_{(k)}\,^\mu}\over {d\tau}}= u^\mu \phi_{(0)(k)}
+\phi_{(k)}\,^{(j)} e_{(j)}\,^\mu\,.
\label{1-2-16}
\end{equation}
Therefore if $\phi_{(j)(k)}=0$, {\it the frame is Fermi-Walker transported
and is non-rotating}. 
It turns out that we can impose $\phi_{(j)(k)}=0$, at least formally.

Suppose that a frame is given such that $\phi_{(j)(k)}\ne 0$. In terms of
the torsion tensor components, the quantities  $\phi_{(j)(k)}$ are written 
as

\begin{equation}
\phi_{(i)(j)}={1\over 2} \lbrack
e_{(i)}\,^\mu e_{(j)}\,^\nu T_{(0)\mu\nu} +
e_{(0)}\,^\mu e_{(j)}\,^\nu T_{(i)\mu\nu} -
e_{(0)}\,^\mu e_{(i)}\,^\nu T_{(j)\mu\nu}\rbrack\,.
\label{1-2-17}
\end{equation}
Under a local Lorentz transformation of the spatial components, we have 

\begin{eqnarray}
\tilde{e}_{(i)}\,^\mu &=& \Lambda_{(i)}\,^{(k)} e_{(k)}\,^\mu \,, 
\label{1-2-18} \\
\tilde{T}_{(i)\mu\nu}&=&
\partial_\mu \tilde{e}_{(i)\nu} -
\partial_\nu \tilde{e}_{(i)\mu} \nonumber \\
{}&=&\Lambda_{(i)}\,^{(k)} T_{(k)\mu\nu}+
\lbrack \partial_\mu \Lambda_{(i)}\,^{(k)}\rbrack e_{(k)\nu}-
\lbrack \partial_\nu \Lambda_{(i)}\,^{(k)}\rbrack e_{(k)\mu}\,.
\label{1-2-19}
\end{eqnarray}
The coefficients $\lbrace \Lambda_{(i)}\,^{(j)}(x) \rbrace$ of the spatial
components of the local Lorentz transformation are fixed by requiring
$\tilde{\phi}_{(i)(j)}=0$.

It is possible to show that 
for given non-vanishing values of the quantities $\phi_{(j)(k)}$, the 
condition $\tilde{\phi}_{(i)(j)}=0$ is obtained provided the coefficients
$\lbrace \Lambda_{(i)}\,^{(j)} \rbrace$ of the Lorentz transformation 
satisfy the equation \cite{Maluf9}

\begin{equation}
e_{(0)}\,^\mu \Lambda^{(j)}\,_{(m)} \partial_\mu \Lambda_{(j)(k)}-
\phi_{(k)(m)}=0\,.
\label{1-2-20}
\end{equation}
Thus, given an arbitrary frame, it is always possible, at least formally, to
rotate the frame and obtain a Fermi-Walker transported frame for which
$\tilde{\phi}_{(i)(j)}=0$. We note that 
the local Lorentz transformation (\ref{1-2-18}) does not affect the timelike 
component $e_{(0)}\,^\mu$.

\subsection{Static frames in the Schwarzschild space-time}

In this and in the following subsections we will analyse some distinguished
frames. We will show that the values of the acceleration tensor are in 
agreement with the physical nature and interpretation of the frames, and have
well defined physical meaning. In these subsections we will adopt the speed 
of light $c=1$.

A frame is static in the Schwarzschild space-time provided it undergoes an 
inertial acceleration that exactly cancels the gravitational acceleration
on the frame. This is an interesting example because it allows an
alternative derivation of the well known gravitational acceleration of a free
particle in the Schwarzschild space-time. 

In spherical coordinates the Schwarzschild space-time is described by the 
line element

\begin{equation}
ds^2=-\biggl(1-{{2m}\over r}\biggr)dt^2 +
\biggl(1-{{2m}\over r}\biggr)^{-1} dr^2+r^2d\theta^2+
r^2\,\sin^2\theta \, d\phi^2\,.
\label{1-3-1}
\end{equation}
A field of static observers in this space-time
is characterized by the vector field $u^\mu$ 
such that $u^\mu=(u^0,0,0,0)$, i.e., the spatial components of $u^\mu$ vanish.
Consequently, in the construction of the tetrad fields we require 

\begin{equation}
e_{(0)}\,^i= u^i=0\,.
\label{1-3-2}
\end{equation}
In view of the orthogonality of the tetrad components, this condition implies
$e^{(k)}\,_0=0$. A simple form of $e_{a\mu}$ in $(t,r,\theta,\phi)$ 
coordinates that satisfies this property and yields Eq. (\ref{1-3-1}) is given
by

\begin{equation}
e_{a\mu}=\pmatrix{-\beta&0&0&0\cr
0&\alpha\sin\theta\,\cos\phi&r\cos\theta\,\cos\phi
&-r\sin\theta\,\sin\phi\cr
0&\alpha\sin\theta\,\sin\phi&r\cos\theta\,\sin\phi
&r\sin\theta\,\cos\phi\cr
0&\alpha\cos\theta&-r\sin\theta&0\cr}\,,
\label{1-3-3}
\end{equation}
where

\begin{eqnarray}
\alpha&=& \biggl(1-{{2m}\over r}\biggr)^{-1/2}\,, \nonumber  \\ 
\beta &=& \biggl(1-{{2m}\over r}\biggr)^{1/2}  \,. 
\label{1-3-4}
\end{eqnarray}
Recall that $a$ and $\mu$ label lines and rows, respectively. 
It is possible to show that in the asymptotic limit 
$r\rightarrow \infty$ the inverse tetrad components in $(t,x,y,z)$
coordinates satisfy 

\begin{eqnarray}
e_{(1)}\,^\mu (t,x,y,z) & \cong &(0,1,0,0)\,, \nonumber \\
e_{(2)}\,^\mu (t,x,y,z) & \cong &(0,0,1,0)\,, \nonumber \\
e_{(3)}\,^\mu (t,x,y,z) & \cong &(0,0,0,1)\,.
\label{1-3-5}
\end{eqnarray}
Altogether, conditions (\ref{1-3-2}) and (\ref{1-3-5}) fix 6 degrees of the
frame.

The evaluation of the acceleration tensor $\phi_{ab}$ is straightforward.
After a number of manipulations we find that Eq. (\ref{1-3-3}) represents 
a non-rotating frame, i.e., 

\begin{equation}
\phi_{(i)(j)}=0\,.
\label{1-3-6}
\end{equation}
The translational acceleration, however, is non-vanishing. From definition 
(\ref{1-10}) we find

\begin{equation}
\phi_{(0)(i)}=T_{(0)(0)(i)} = 
e_{(0)}\,^\mu e_{(i)}\,^\nu T_{(0)\mu\nu}\,.
\label{1-3-7}
\end{equation}
For $a=(0)$, the only non-vanishing component of $T_{a\mu\nu}$ is
$T_{(0)01}=\partial_1 \beta$. Thus, the equation above yields

\begin{equation}
\phi_{(0)(i)}= g^{00}g^{11}e_{(0)0} e_{(i)1} T_{(0)01}\,,
\label{1-3-8}
\end{equation}
from what follows

\begin{eqnarray}
\phi_{(0)(1)}&=&{m\over r^2}\biggl( 1-{{2m}\over r}\biggr)^{-1/2}
\sin\theta\,\cos\phi \,,\nonumber \\
\phi_{(0)(2)}&=&{m\over r^2}\biggl( 1-{{2m}\over r}\biggr)^{-1/2}
\sin\theta\,\sin\phi \,,\nonumber \\
\phi_{(0)(3)}&=&{m\over r^2}\biggl( 1-{{2m}\over r}\biggr)^{-1/2}
\cos\theta \,.
\label{1-3-9}
\end{eqnarray}
We define the acceleration ${\bf a}$,

\begin{equation}
{\bf a}= \phi_{(0)(1)} \hat{{\bf x}}+\phi_{(0)(2)} \hat{{\bf y}}
+\phi_{(0)(3)} \hat{{\bf z}}\,,
\label{1-3-10}
\end{equation}
which may be written as

\begin{equation}
{\bf a}= {m\over r^2}\biggl( 1-{{2m}\over r}\biggr)^{-1/2}\,
\hat{\bf r}\,,
\label{1-3-11}
\end{equation}
where

$${\hat{\bf r}}=\sin\theta\,\cos\phi\,\hat{{\bf x}} +
\sin\theta\,\sin\phi\,\hat{{\bf y}} +\cos\theta\,\hat{{\bf z}}\,. $$

Equation (\ref{1-3-11}) represents the inertial acceleration necessary to
maintain the frame in stationary state in space-time. Therefore it must
necessarily cancel the geodesic gravitational acceleration that is exerted
on the frame. In fact,

\begin{equation}
a= -{m\over r^2}\biggl( 1-{{2m}\over r}\biggr)^{-1/2}\,,
\label{1-3-12}
\end{equation}
is precisely the standard geodesic acceleration of a body in free fall in the
Schwarzschild space-time, which may be obtained directly from Eq. (\ref{1-1})
\cite{Hartle}. 

We have seen that Eqs. (\ref{1-3-2}) and (\ref{1-3-5}) fix the
six degrees of freedom of the tetrad frame (\ref{1-3-3}). Alternatively, the
set of conditions (\ref{1-3-6}), (\ref{1-3-9}) also 
fix the frame. The sets (\ref{1-3-6}), (\ref{1-3-9}) and (\ref{1-3-2}), 
(\ref{1-3-5}) are essentially equivalent. This equivalence is in agreement 
with the discussion presented in the paragraph just below Eq. (\ref{1-10}),
namely, either we establish the frame by fixing the frame components
according to the symmetry of the physical configuration, and which amounts to
choosing a suitable congruence of timelike worldlines, or by fixing the
components of the acceleration tensor (which, in general, is a less 
intuitive procedure). We remark, finally, that Eq. (\ref{1-3-7}) is 
invariant under coordinate transformations. 

\subsection{Frame in free fall in the Schwarzschild space-time}

A frame in free fall in the Schwarzschild space-time is radially accelerated
towards the center of the black hole, under the action of the gravitational
field only. 

An observer that is in radial free fall in the Schwarzschild space-time is 
endowed with the four-velocity \cite{Hartle}

\begin{equation}
u^\alpha=
\biggr[ \biggl(1-{{2m}\over r}\biggr)^{-1}, 
-\biggl({{2m}\over r}\biggr)^{1/2},
0,0\biggr]\,.
\label{1-4-1}
\end{equation}
The simplest set of tetrad fields that satisfies the condition

\begin{equation}
e_{(0)}\,^\alpha=u^\alpha\,,
\label{1-4-2}
\end{equation}
is given by \cite{Maluf8}

\begin{equation}
e_{a\mu}=\pmatrix{-1&-\alpha^2 \eta&0&0\cr
\eta \sin\theta \cos\phi & \alpha^2 \sin\theta \cos\phi&
r \cos\theta \cos\phi & -r \sin\theta \sin\phi\cr
\eta \sin\theta \sin\phi & \alpha^2 \sin\theta \sin\phi&
r \cos\theta \sin\phi &  r \sin\theta \cos\phi\cr
\eta \cos\theta & \alpha^2 \cos\theta & -r\sin\theta&0}\,,
\label{1-4-3}
\end{equation}
where
\begin{equation}
\eta=\biggl({{2m}\over r}\biggr)^{1/2}=(1-\alpha^{-2})^{1/2}\,.
\label{1-4-4}
\end{equation}

Since the frame is in free fall, the equation 
$\phi_{(0)(i)}=0$ is verified.
It is not difficult to show by direct calculations  that this set of tetrad
fields also satisfies the conditions 

\begin{equation}
\phi_{(i)(j)}=
{1\over 2}\lbrack T_{(0)(i)(j)}+T_{(i)(0)(j)}-T_{(j)(0)(i)}
\rbrack=0\,.
\label{1-4-5} 
\end{equation}

Equation (\ref{1-4-2}) fixes 3 conditions on the frame, i.e., 
$e_{(0)}\,^i=u^i$. Together with the 3
conditions fixed by Eq. (\ref{1-4-5}), we have six conditions on the frame.
These conditions completely fix the structure of the tetrad fields, even 
though Eq. (\ref{1-4-5}) has been verified {\it a posteriori}.
Therefore Eq. (\ref{1-4-3}) describes
a non-rotating frame in radial free fall in the Schwarzschild space-time. 
No inertial acceleration is imparted to the frame.

\subsection{Static frames in the Kerr space-time}

Another interesting physical configuration is the frame adapted to static 
observers in the Kerr space-time. In the weak field approximation, a 
relationship may be established with the gravitoelectromagnetic (GEM) field 
quantities \cite{Maluf8}. In spherical (Boyer-Lindquist) coordinates, the 
Kerr's space-time is established by the line element

\begin{eqnarray}
ds^2&=&
-{{\psi^2}\over {\rho^2}}dt^2-{{2\chi\sin^2\theta}\over{\rho^2}}
\,d\phi\,dt
+{{\rho^2}\over {\Delta}}dr^2 \nonumber \\
&{}&+\rho^2d\theta^2+ {{\Sigma^2\sin^2\theta}\over{\rho^2}}d\phi^2\,,
\label{1-5-1}
\end{eqnarray}
with the following definitions:

\begin{eqnarray}
\Delta&=& r^2+a^2-2mr\,, \nonumber \\
\rho^2&=& r^2+a^2\cos^2\theta \,,  \nonumber \\
\Sigma^2&=&(r^2+a^2)^2-\Delta a^2\sin^2\theta\,, \nonumber \\
\psi^2&=&\Delta - a^2 \sin^2\theta\,, \nonumber \\
\chi &=&2amr\,.
\label{1-5-2}
\end{eqnarray}

A static frame in Kerr's space-time is defined by the congruence of timelike
curves $u^\mu(\tau)$ such that $u^i=0$, namely, the spatial velocity of the 
observers is zero with respect to static observers at spacelike infinity. 
Since we identify $e_{(0)}\,^i=u^i$, a static reference frame is established
by the condition

\begin{equation}
e_{(0)}\,^i=0\,.
\label{1-5-3}
\end{equation}
As in subsection 2.2, 
the orthogonality of the tetrads imply 
$e^{(k)}\,_0=0$. This latter equation remains valid even after a local 
rotation of the frame:
$\tilde e^{(k)}\,_0=\Lambda^{(k)}\,_{(j)} e^{(j)}\,_0=0$.
Therefore, condition (\ref{1-5-3}) determines the static character of the
frame, up to an orientation of the spacelike components of the frame in the
three-dimensional space (see Eq. (\ref{1-5-8}) below). 

A simple form of the tetrad fields that satisfies Eq. (\ref{1-5-3}) (or, 
equivalently, $e^{(k)}\,_0=0$) reads \cite{Maluf8}

\begin{equation}
e_{a\mu}=\pmatrix{-A&0&0&-B\cr
0&C\sin\theta\cos\phi& \rho\cos\theta\cos\phi&-D\sin\theta\sin\phi\cr
0&C\sin\theta\sin\phi& \rho\cos\theta\sin\phi&D\sin\theta\cos\phi\cr
0&C\cos\theta&-\rho\sin\theta&0}\,,
\label{1-5-4}
\end{equation}
with the following definitions,

\begin{eqnarray}
A&=& {\psi \over \rho}\,, \nonumber \\
B&=& {{\chi \sin^2\theta}\over {\rho \psi}}\,, \nonumber \\
C&=&{\rho \over \sqrt{\Delta}}\,, \nonumber \\
D&=& {\Upsilon \over{\rho \psi}}\,.
\label{1-5-5}
\end{eqnarray}
In the expression of $D$ we have

\begin{equation}
\Upsilon =(\psi^2\Sigma^2+\chi^2\sin^2\theta)^{1/2}\,.
\label{1-5-6}
\end{equation}

We are interested in obtaining the acceleration tensor $\phi_{ab}$,
and for this purpose we need the inverse tetrad fields  $e_a\,^\mu$. 
They read

\begin{equation}
e_a\,^\mu=\pmatrix{ {\rho \over \psi} & 
{{\rho \chi}\over{\psi \Upsilon}}  \sin\theta\sin\phi &
-{{\rho \chi}\over{\psi \Upsilon}} \sin\theta\cos\phi & 0\cr
0 & {{\sqrt{\Delta}}\over \rho}\sin\theta \cos\phi &
{{\sqrt{\Delta}}\over \rho}\sin\theta \sin\phi &
{{\sqrt{\Delta}}\over \rho} \cos\theta \cr
0 & {1\over \rho}\cos\theta \cos\phi &
{1\over \rho}\cos\theta \sin\phi & -{1\over \rho} \sin\theta \cr
0 & -{{\rho \psi}\over {\Upsilon}} {{\sin\phi}\over{\sin\theta}} &
{{\rho \psi}\over {\Upsilon}} {{\cos\phi}\over{\sin\theta}} & 0}\,,
\label{1-5-7}
\end{equation}
where now the index $a$ labels the columns and $\mu$ labels the rows.

The frame determined by Eqs. (\ref{1-5-4}) and (\ref{1-5-5}) is valid in the
region outside the ergosphere. The function $\psi^2=\Delta - a^2 \sin^2\theta$
vanishes over the external surface of the ergosphere (defined by 
$r=r^{\star}=m+\sqrt{m^2-a^2\cos^2\theta}\,$; over this surface we have
$g_{00}=0$). Various components of Eqs. (\ref{1-5-4}) and (\ref{1-5-5}) are 
not well defined when $r=r^{\star}$. It is well known that it is not possible 
to maintain static observers inside the ergosphere of the Kerr space-time.

It follows from Eq. (\ref{1-5-7}) that for large values of $r$ we have

$$ e_{(3)}\,^\mu (t,r,\theta,\phi)
\cong (0,\cos\theta, -(1/r)\sin\theta,0)\,, $$
or

\begin{equation}
e_{(3)}\,^\mu (t,x,y,z)
\cong (0,0,0,1)\,.
\label{1-5-8}
\end{equation}
Therefore we may assert that the frame given by Eq. (\ref{1-5-4}) is
characterized by the following properties: {\bf (i)} the frame is static, 
because Eq. (\ref{1-5-3}) is verified; {\bf (ii)} the $e_{(3)}\,^\mu$
components are oriented along the symmetry axis of the black hole (the $z$ 
direction). The second condition is ultimately reponsible for the simple 
form of Eq. (\ref{1-5-4}).

The evaluation of $\phi_{ab}$ requires more algebraic manipulations than the
previous cases, but is straightforward. It is useful to define the vectors

\begin{eqnarray}
{\bf \hat{r}}&=& \sin\theta \cos\phi\,{\bf \hat{x}}+
\sin\theta \sin\phi\,{\bf \hat{y}}+
\cos\theta \,{\bf \hat{z}}\,, \nonumber \\
{\bf \hat{\theta}}&=&\cos\theta \cos\phi\,{\bf \hat{x}}+
\cos\theta \sin\phi\,{\bf \hat{y}}-
\sin\theta\, {\bf \hat{z}}\,,
\label{1-5-9}
\end{eqnarray}
which have well defined meaning as unit vectors in the asymptotic
limit $r \rightarrow \infty$. The translational acceleration and the 
frequency of rotation are denoted in Cartesian components as  

\begin{eqnarray}
{\bf a}&=& (\phi_{(0)(1)}, \phi_{(0)(2)}, \phi_{(0)(3)})\,, 
\label{1-5-10} \\
{\bf \Omega}&=& (\phi_{(2)(3)}, \phi_{(3)(1)}, \phi_{(1)(2)})\,.
\label{1-5-11}
\end{eqnarray}
We obtain the following expressions for ${\bf a}$ and ${\bf \Omega}$
\cite{Maluf8}:

\begin{eqnarray}
{\bf a}&=& {m \over \psi^2}\biggl[
{{\sqrt{\Delta} \over {\rho}}\biggl(
{{2r^2}\over \rho^2}-1 \biggr){\bf \hat{r}}\, +
{{2ra^2}\over {\rho}^3}}\sin\theta \cos\theta\,{\bf \hat{\theta}}
\biggr]\,,
\label{1-5-12} \\
{\bf \Omega}&=& -{\chi \over{\Upsilon \rho }}\cos\theta\,{\bf \hat{r}}+
{{\psi^2 \sqrt{\Delta}}\over {2\Upsilon \rho}}\sin\theta\,
\partial_r\biggl( {\chi \over {\psi^2}}\biggr)\, {\bf \hat{\theta}}
-{{\psi^2} \over {2\Upsilon \rho}}\sin\theta\,\partial_\theta 
\biggl({\chi \over {\psi^2}} \biggr) {\bf \hat{r}}\,.
\label{1-5-13}
\end{eqnarray}

These vectors also characterize the frame determined by Eq. (\ref{1-5-4}). The
fixation of {\bf a} and ${\bf \Omega}$ is equivalent to the fixation of
six components of the tetrad fields. Equations (\ref{1-5-12}) and 
(\ref{1-5-13}) represent the inertial accelerations that one must exert on the
frame in order to ensure {\bf (i)} the static character of the frame
(Eq. (\ref{1-5-3})), and that {\bf (ii)} the $e_{(3)}\,^\mu$ components of the 
tetrad fields asymptotically coincide with the symmetry axis of the black hole
(Eq. (\ref{1-5-8})).

The form of {\bf a} and ${\bf \Omega}$ for large values of $r$ is very 
interesting. It is easy to verify that in the limit $r \rightarrow \infty$ we 
obtain

\begin{eqnarray}
{\bf a} & \cong & {m \over r^2}\,{\bf \hat{r}}\,, 
\label{1-5-14} \\
{\bf \Omega} & \cong & -{{am} \over r^3}
\biggl(2\cos\theta\, {\bf \hat{r}}+ \sin\theta\, {\bf \hat{\theta}}
\biggr)\,.
\label{1-5-15}
\end{eqnarray}
By identifying $ m \leftrightarrow q$ and $4\pi ma \leftrightarrow \bar{m}$,
where $q$ is the electric charge and $\bar m$ is the magnetic dipole moment, 
equations (\ref{1-5-14}) and (\ref{1-5-15}) resemble the electric field of a 
point charge and the magnetic field of a perfect dipole that points in the 
$z$ direction, respectively. These equations represent a manifestation of 
gravitoelectromagnetism.

If we remove all inertial accelerations on the frame, an observer located at
a position $(r,\theta,\phi)$ will be subject to a radial acceleration 
$-{\bf a}$ and to a frequency of rotation determined by 
$-{\bf \Omega}={\bf \Omega}_D$,
which is the dragging frequency of the frame. Thus, the gravitomagnetic 
effect is locally equivalent to inertial effects in a frame endowed with 
angular frequency $-{\bf \Omega}_D$. This is precisely the gravitational 
Larmor's theorem, discussed in Ref. \cite{Mashh1}.\par 
\bigskip

In view of the standard examples discussed above, it is quite clear that the
primary and natural attribute of tetrad fields is the establishment of 
reference frames in space-time. The inevitable conclusion is that tetrad 
fields carry information of both the inertial state of the observer and of 
the gravitational field.

\section{The gravitational energy-momentum definitions in the TEGR}

In the ordinary metric formulation of general relativity, the gravitational
energy-momentum was first given by the standard ADM expressions \cite{ADM},
but the complete set of energy, momentum, angular momentum and centre of 
mass moment was probably put forward 
in a concise form by Regge and Teitelboim \cite{Regge}. 
These notions were presented in the context of the Hamiltonian formulation of
general relativity. The idea was to require the variation of the total 
Hamiltonian to be well defined in an asymptotically flat space-time, where 
the standard asymptotic space-time translations and 4-rotations are 
considered as coordinate transformations at spacelike infinity. 
This requirement leads to the addition of boundary (surface) terms
to the primary Hamiltonian, so that the latter has well defined functional
derivatives, and therefore one may obtain the field equations in the 
Hamiltonian framework (Hamilton's equations) by means of a consistent 
procedure.  In this way, one arrives at the total energy, momentum, angular
momentum and centre of mass moment of the gravitational field, given by 
surface terms of the total Hamiltonian. The work of Regge and Teitelboim 
indicate the relevance of the Hamiltonian formulation for the establishment 
of these definitions. 

In this Section, we will recall the definitions of energy-momentum $P^a$ 
and of the 4-angular momentum $L^{ab}$ for the gravitational field in the 
TEGR. These definitions have been discussed several times in the literature.
Here, we will display the summary presented in Refs. \cite{Maluf3,Maluf11}. 
In the present review, the TEGR is constructed out of the tetrad fields only,
as we pointed out earlier. In the last Section, we 
will briefly discuss the so called ``covariant" formulation, which
encompasses the addition of a flat spin connection. 

The first relevant consideration is an identity between the scalar curvature 
of the Riemann-Christoffel tensor, 
and an invariant combination of quadratic terms of the torsion tensor, 

\begin{equation}
eR(e) \equiv -e\left({1\over 4}T^{abc}T_{abc} + 
{1\over 2}T^{abc}T_{bac} - T^{a}T_{a}\right)
+ 2\partial_{\mu}(eT^{\mu})\,,
\label{2}
\end{equation}
where $ T_{a} = T^{b}\,_{ba}$ and 
$T_{abc} = e_{b}\,^{\mu}e_{c}\,^{\nu}T_{a\mu\nu}$.
The Lagrangian density for the gravitational field in the TEGR is given by 
\cite{Maluf1}

\begin{eqnarray}
L(e) &=& -k\,e\left({1\over 4}T^{abc}T_{abc} + {1\over 2}T^{abc}T_{bac} 
- T^{a}T_{a}\right) - {1\over c}L_{M}\nonumber \\
& \equiv & -ke\Sigma^{abc}T_{abc} - {1\over c}L_{M}\,,
\label{3}
\end{eqnarray}
where $k = c^3/(16\pi G)$, $L_{M}$ represents the Lagrangian density for the
matter fields, and $\Sigma^{abc}$ is defined by
\begin{equation}
\Sigma^{abc} = {1\over 4}\left(T^{abc} + T^{bac} - T^{cab}\right) 
+ {1\over 2}\left(\eta^{ac}T^{b} - \eta^{ab}T^{c}\right)\,.
\label{4}
\end{equation}
Thus, the Lagrangian density is geometrically equivalent to the scalar 
curvature density. The variation of $L(e)$ with respect to $e^{a\mu}$
yields the fields equations 

\begin{equation}
e_{a\lambda}e_{b\mu}\partial_\nu (e\Sigma^{b\lambda \nu} )-
e (\Sigma^{b\nu}\,_aT_{b\nu\mu}-
{1\over 4}e_{a\mu}T_{bcd}\Sigma^{bcd} )={1\over {4kc}}e\texttt{T}_{a\mu}\,,
\label{5-1}
\end{equation}
where $\texttt{T}_{a\mu}$ is defined by 
${{\delta L_M}/ {\delta e^{a\mu}}}=e\texttt{T}_{a\mu} $.
These field equations are equivalent to Einstein's equations. It is possible
to verify by direct calculations that the equations above can be rewritten
as 

\begin{equation}
\label{5-2}
{1\over 2}\lbrack R_{a\mu}(e) - {1\over 2}e_{a\mu}R(e)\rbrack
={1\over {4kc}}\,\texttt{T}_{a\mu}\,,
\end{equation}

The left and right hand sides of Eq. (\ref{2}) are invariant under 
arbitrary SO(3,1) 
local transformations. However, the Lagrangian density (\ref{3})
does not contain the total divergence that appears on the right hand side of
(\ref{2}). Therefore, the Lagrangian density (\ref{3}) is not invariant
under local Lorentz transformations, but the field equations are covariant
under such transformations. As discussed in Ref. \cite{Maluf1}, the variation
of the action integral constructed out of (\ref{3}) is well defined if one 
considers asymptotically flat space-times, i.e., one does not need the 
addition of surface terms to obtain well defined variations in the asymptotic
limit $r\rightarrow \infty$.

Equations (\ref{5-1}) and (\ref{5-2}) prove the equivalence between the
TEGR and the standard metric formulation of GR. Both theories are equivalent
because the field equations are the same. However, in a theory constructed
out of tetrad fields, additional field quantities such as third rank tensors
do exist (the torsion tensor, for instance), 
besides new vector and tensor densities, which cannot be defined in the
standard metric formulation of gravity. These additional field quantities are
covariant under global Lorentz transformations, but not under local  
transformations. It turns out that these vector and tensor densities yield
the definitions of energy, momentum and 4-angular momentum of the 
gravitational field, which are covariant under global Lorentz 
transformations, as expected. This general framework is not different from
what happens in the ordinary formulation of arbitrary field theories, where
the energy-momentum and 4-angular momentum are frame dependent field 
quantities that transform under the global Lorentz group. Thus, energy 
transforms as the zero component of the energy-momentum four-vector. These
features must naturally hold also in the context of the theory
for the gravitational field.
Consider a static black hole represented by a mass parameter $m$ only. For a
distant observer, the total energy of this black hole is given by $E=mc^2$.
But at very great distances, this black hole is considered as a particle of
mass $m$, and if it moves with constant velocity $v$, then its total energy
as seen by the same distant observer is $E=\gamma mc^2$, where 
$\gamma = (1-v^2/c^2)^{-1/2}$. By the same reasoning, 
the gravitational momentum, angular momentum and the 
centre of mass moment are also frame dependent field quantities, 
whose values are different in different frames and for different observers. 
On physical grounds, energy, momentum, angular momentum and centre of mass 
moment cannot be local Lorentz {\it invariant} field quantities, since these 
quantities depend on the frame, as we know from special
relativity, which is the limit of the general theory of relativity when the 
gravitational field is weak or negligible.

After some rearrangements, Eq. (\ref{5-1}) may be written in the form 
\cite{Maluf10}

\begin{equation}
\partial_{\nu}(e\Sigma^{a\mu\nu}) = 
{1\over 4k}ee^{a}\,_{\nu}(t^{\mu\nu} + {1\over c}\texttt{T}^{\mu\nu})\,,
\label{5}
\end{equation}
where 
\begin{equation}
t^{\mu\nu} = k(4\Sigma^{bc\mu}T_{bc}\,^{\nu} - 
g^{\mu\nu}\Sigma^{bcd}T_{bcd})\,,
\label{6}
\end{equation}
is interpreted as the gravitational energy-momentum tensor
\cite{Maluf3,Maluf10}, and $\texttt{T}^{\mu\nu} = 
e_{a}\,^{\mu}\texttt{T}^{a\nu}$.

\subsection{Main results from the Hamiltonian formulation}

In this subsection we will consider the field equations in the absence of
matter fields.
The Hamiltonian density of the TEGR is constructed as usual in the phase 
space of the theory. We first note that the Lagrangian density (\ref{3}) 
does not depend on the time derivatives of $e_{a0}$. Therefore, the latter 
arise as Lagrange multipliers in the Hamiltonian density $H$. The momenta 
canonically conjugated to $e_{a0}$ are denoted by $\Pi^{a0}$. The latter are 
primary constraints of the
theory: $\Pi^{a0} \approx 0$. The momenta canonically conjugated to $e_{ai}$
are given by $\Pi^{ai} = \delta L/\delta \dot{e}_{ai}= -4k\Sigma^{a0i}$.
The Hamiltonian density $H$ is obtained by rewriting the 
Lagrangian density in the form $L = \Pi^{ai}\dot{e}_{ai} - H$,
in terms of $e_{ai}, \Pi^{ai}$ and Lagrange multipliers. After the Legendre
transform is performed, we obtain the final form of the Hamiltonian density.
It reads \cite{Maluf12,Maluf13}

\begin{equation}
H(e,\Pi) = e_{a0}C^{a} + \lambda_{ab}\Gamma^{ab},
\label{7}
\end{equation}
where $\lambda_{ab}$ are Lagrange multipliers. In the equation above we have
omitted a surface term. The expression of $C^{a} = \delta H/\delta e_{a0}$
in terms of the field variables is rather long. It is formally 
simplified in Eq. (\ref{9}) below. The constraints $\Gamma^{ab}=-\Gamma^{ba}$
are defined by

\begin{equation}
\Gamma^{ab} = 2\Pi^{[ab]} + 4ke(\Sigma^{a0b} - \Sigma^{b0a})\,.
\label{8}
\end{equation}
After solving the field equations, the Lagrange multipliers are identified as
$\lambda_{ab} = (1/4)(T_{a0b}-T_{b0a}+e_{a}\,^{0}T_{00b}-e_{b}\,^{0}T_{00a})$
\cite{Maluf12}. The constraints $C^{a}$ may be written as

\begin{equation}
C^{a} = -\partial_{i}\Pi^{ai} - p^{a} = 0\,,
\label{9}
\end{equation}
where $p^{a}$ is an intricate expression of the field quantities. In the 
context of the Lagrangian density (\ref{3}), $p^a$ may be interpreted as the
gravitational energy-momentum density (see Eq. (\ref{10}) below). 
The quantities
$C^{a}$ and $\Gamma^{ab}$ are first class constraints. They satisfy an 
algebra  similar to the algebra of the Poincar\'e group \cite{Maluf13}. 

The integral form of the constraint equations $C^{a} = 0$ yields the 
gravitational energy-momentum $P^{a}$ \cite{Maluf10,Maluf14},

\begin{equation}
P^{a} = - \int_{V}d^{3}x\,\partial_{i}\Pi^{ai}\,,
\label{10}
\end{equation}
where $V$ is an arbitrary volume of the three-dimensional space and 
$\Pi^{ai} = -4k\Sigma^{a0i}$. By means of Gauss's law, we may transform the
volume integral in Eq. (\ref{10}) into a surface integral, according to

\begin{equation}
\label{11}
P^a=\oint_S dS_i\,4k\Sigma^{a0i}\,,
\end{equation}
where $S$ is the 2-dimensional spacelike boundary of the volume $V$.

In similarity to the definition above, the 
definition of the gravitational 4-angular momentum follows from the 
integral form of the constraint equations $\Gamma^{ab}=0$ \cite{Maluf15}. 
It reads

\begin{equation}
L^{ab} = -\int_{V}d^{3}x\,2\Pi^{[ab]}\,,
\label{12}
\end{equation}
where 
\begin{equation}
2\Pi^{[ab]} =(\Pi^{ab} - \Pi^{ba}) =
-4ke(\Sigma^{a0b} - \Sigma^{b0a})\,.
\label{13}
\end{equation}

However, it has been 
noted \cite{Rocha} that the second term on the right hand side of 
Eq. (\ref{8}) can be rewritten as a total divergence, so that the constraints
$\Gamma^{ab}$ become

\begin{equation}
\Gamma^{ab} = 2\Pi^{[ab]} - 2k\partial_{i}[e(e^{ai}e^{b0} -
 e^{bi}e^{a0})] = 0\,.
\label{14}
\end{equation}
Therefore, the definition of the total 4-angular momentum of the gravitational
field $L^{ab}$ may be given by an integral of a total divergence, in 
similarity to Eq. (\ref{10}),

\begin{equation}
\label{15}
L^{ab}=\oint_S dS_i\, 2k[e(e^{ai}e^{b0} - e^{bi}e^{a0})]\,.
\end{equation}

It is easy to show \cite{Maluf15} that expressions (\ref{10}) and (\ref{12}) 
satisfy the algebra of the Poincar\'e group in the phase space of the theory,

\begin{eqnarray}
\{P^{a}, P^{b}\}& =&0\,, \nonumber \\
\{P^{a},L^{bc}\}&=& \eta^{ab}P^{c} - \eta^{ac}P^{b}\,,  \nonumber \\
\{L^{ab},L^{cd}\} &=& \eta^{ad}L^{cb} + \eta^{bd}L^{ac} - 
\eta^{ac}L^{db} - \eta^{bc}L^{ad}\,.
\label{16}
\end{eqnarray}
The quantities in the left hand side of the expressions above are Poisson 
brackets evaluated in the phase space determined by the canonically conjugate
field variables ($e_{ai}$, $\Pi^{ ai}$). In the evaluation of the Poisson
brackets, we have used the following 
functional derivatives of $P^a$ and $L^{ab}$, 

\begin{eqnarray}
{{\delta L^{ab}}\over {\delta e_{ck}(z)}}&=&
-\eta^{ac} \Pi^{bk}(z) + \eta^{bc} \Pi^{ak}(z)\,, 
\nonumber \\
{{\delta L^{ab}}\over {\delta \Pi_{ck}(z)}}&=&
\delta^a_c e^b\,_k(z)- \delta^b_c e^a\,_k(z)\,,
\nonumber \\
{{\delta P^a} \over {\delta e_{ck}(z)}}&=&0\,, \nonumber \\
{{\delta P^a} \over {\delta \Pi_{ck}(z)}}&=&-
\int d^3x \delta^a_c {\partial \over {\partial x^k}}
\delta^3(x-z)\,.
\label{4-4-1}
\end{eqnarray}

Therefore, from a physical point of view, the interpretation of the 
quantities $P^{a}$ and $L^{ab}$ as gravitational energy-momentum and 
4-angular momentum is consistent. It is well known that the algebra of the
Poincar\'e group is intimately connected to energy, momentum and 4-angular
momentum. The {\it values} of $P^a$ and $L^{ab}$ may
be computed by means of Eqs. (\ref{11}) and (\ref{15}). 

Definitions (\ref{10}) and (\ref{12}) are invariant under coordinate 
transformations of the three-dimensional space, under time
reparametrizations, and under global SO(3,1) transformations. The 
gravitational energy is the zero component of the energy-momentum 
four-vector $P^a$. 

\subsection{Gravitational energy-momentum from the Lagrangian field equations}

The definition of gravitational energy-momentum may also be obtained in the
Lagrangian framework, according to the procedure of Ref. \cite{Maluf10}. 
Again, we assume $c=1=G$, and consider the field equations in the form given
Eqs. (\ref{5}) and (\ref{6}).
In view of the antisymmetry property 
$\Sigma^{a\mu\nu}=-\Sigma^{a\nu\mu}$, it follows that

\begin{equation}
\partial_\lambda
\left[e\, e^a\,_\mu( t^{\lambda \mu} + \texttt{T}^{\lambda \mu})\right]=0\,.
\label{17}
\end{equation}
The equation above yields the continuity (or balance) equation,

\begin{equation}
{d\over {dt}} \int_V d^3x\,e\,e^a\,_\mu (t^{0\mu} +\texttt{T}^{0\mu})
=-\oint_S dS_j\,
\left[e\,e^a\,_\mu (t^{j\mu} +\texttt{T}^{j\mu})\right]\,,
\label{18}
\end{equation}
where $S$ is the boundary of an arbitrary 3-dimensional volume $V$, as 
considered above. Therefore we (again) identify
$t^{\lambda\mu}$ as the gravitational energy-momentum tensor \cite{Maluf10},

\begin{equation}
P^a=\int_V d^3x\,e\,e^a\,_\mu (t^{0\mu} 
+\texttt{T}^{0\mu})\,,
\label{19}
\end{equation}
as the {\it total} energy-momentum contained within the volume $V$ of the 
3-dimensional space,

\begin{equation}
\Phi^a_g=\oint_S dS_j\,
\, (e\,e^a\,_\mu t^{j\mu})\,,
\label{20}
\end{equation}
as the gravitational energy-momentum flux, and

\begin{equation}
\Phi^a_m=\oint_S dS_j\,
\,( e\,e^a\,_\mu \texttt{T}^{j\mu})\,,
\label{21}
\end{equation}
as the energy-momentum flux of matter \cite{Maluf10}. In view of the field
equations (\ref{5}), Eq. (\ref{19}) may be written as 
$P^a=-\int_V d^3x \partial_j \Pi^{aj}\,,$
from what follows

\begin{equation}
P^a=-\oint_S dS_j\,\Pi^{aj}\,,
\label{22}
\end{equation}
where $\Pi^{aj}=-4ke\,\Sigma^{a0j}$, in similarity to Eq. (\ref{11}).

We establish Eqs. (\ref{11}) and (\ref{15}) as the definitions of the
gravitational energy-momentum and gravitational 4-angular momentum,
respectively. The reason for considering surface integrals
is because the gravitational field on the
surface of integration $S$ contains all the information on the interior 
region, and of course the integral can be carried out more easily, even in
the presence of singularities, admitting that the space-time has 
singularities. We cannot transform the volume integral (\ref{10}) into the
surface integral (\ref{11}) by means of Gauss theorem, if the space-time
has singularities, and for this reason Eq. (\ref{11}) is conceptually more
important than Eq. (\ref{10}). But we recall that exactly this same 
difficulty occurs in the definition of the ADM gravitational energy-momentum,
whose definitions are given by surface integrals and yield satisfactory and
acceptable results in the context of asymptotically flat space-times. 
Therefore, Eq. (\ref{11}) represents the gravitational energy-momentum within
the surface $S$.

Definition (\ref{11}) (or (\ref{22})) is invariant under coordinate 
transformations of the three-dimensional space and under time 
reparametrizations, which are basic requirements of any gravitational
energy-momentum definition. Eq. (\ref{17}) is a true energy-momentum 
conservation equation, in contrast to the analogous situation in the
metric formulation of general relativity, where the covariant derivative of
the energy-momentum tensor is not a strict conservation equation. We repeat
and emphasize that in
the ordinary formulation of arbitrary field theories, energy, 
momentum, angular momentum and the centre of mass
moment are frame dependent field quantities, that 
transform under the global SO(3,1) group. This property, which takes place
in special relativity, is satisfied by 
definition (\ref{11}). In particular, the energy $E$  
transforms as the zero component of the energy-momentum four-vector. Thus,
we conclude that in the context of the TEGR, there is a smooth transition
between the weak field limit of general relativity and special relativity.
We have a well defined conservation equation for the matter fields, with or
without the gravitational field, in arbitrary coordinate systems.

The components of the vector $P^a$ are $({E/c}, {\bf P})$.
If we assume that the tetrad fields satisfy asymptotic boundary conditions,

\begin{equation}
e_{a\mu} \simeq \eta_{a\mu}
+ {1\over 2}h_{a\mu}(1/r)\,,
\label{4-2-4}
\end{equation}
at spatial infinity, i.e., in the limit $r \rightarrow \infty$, then the 
total gravitational energy $E=cP^{(0)}$ is the ADM energy, 

\begin{equation}
E ={{c^4}\over {16\pi G}}\int_{S\rightarrow \infty}dS_k(\partial_i
h_{ik}-\partial_k h_{ii}) = E_{ADM}\,.
\label{4-2-5}
\end{equation}
The expression above is just one of the consistent results that follow from
the above definitions.

\section{Gravitational energy in the Kerr space-time}

The Kerr space-time plays a central role in general relativity. We may assume
that practically all single, self gravitating systems in nature rotate around
their own symmetry or principal inertia 
axes, and that the Kerr space-time represents the exterior region
of these rotating compact objects. However, the Kerr line element represents
a black hole and possesses the ergoregion, which is the region between the
external event horizon ($r=r_+$, see below) and the ergosurface 
($r=r_+^\star$). It is
theoretically predicted that it is possible to extract energy from the region
between $r=r_+^\star$ and $r=r_+$ by means of the Penrose process 
\cite{Penrose}. After the complete extraction of this energy, which is 
considered to be a rotational energy, when the black
hole no longer rotates, there remains
the irreducible mass of the black hole $M_{irr}$, which was calculated by 
Christodoulou and Ruffini \cite{CR}. It is also assumed that no form of 
energy can escape from the external event horizon of the black hole,
during the complete Penrose process. For this reason, the evaluation of the 
gravitational energy contained within the external event horizon of the Kerr
black hole is a relevant issue in the investigation of the localization of 
the gravitational energy. It is rather obvious that this energy cannot be 
made to vanish by means of coordinate transformations, or by any formulation
of the principle of equivalence, not even at a point inside the external 
event horizon.

In order to calculate the energy contained within the external event horizon
of the Kerr black hole, we need tetrad fields that are defined up to $r=r_+$,
i.e., we must have observers that can reach the external event horizon,
even under the dragging effect of the black hole, since inside the ergophere 
it is not possible to establish static frames.
The tetrad fields considered in subsection 2.4 do not satisfy this 
requirement, since they are defined from $r=\infty$ up to $r=r_+^\star$, 
which defines the external surface of the ergosphere. The
set of tetrad fields that are defined from $r=\infty$ up to $r=r_+$ is the 
one that satisfies Schwinger's time gauge condition \cite{Schwinger}.

We will make use of the notation in Section 2.4, and in particular of Eqs.
(\ref{1-5-1}) and (\ref{1-5-2}), but not of Eq. (\ref{1-5-3}), as we will
explain ahead. The Kerr black hole is characterized by the mass parameter 
$m$ and by the angular momentum per unit mass $a=J/m$. 
In the Penrose process \cite{Penrose}, the initial mass $m$ and 
angular momentum $J$ of the black hole vary by $dm$ and $dJ$, respectively, 
such that $dm -\Omega_H dJ \ge 0$, where $\Omega_H$ is the angular velocity 
of the external event horizon of the black hole,

\begin{equation}
\Omega_H={a\over {2mr_+}} ={a \over {a^2+r_+^2}}\,.
\label{5-1-3}
\end{equation}
The radius of the external event horizon is given by $r_+=m+\sqrt{m^2-a^2}$.
During the process, the variation of the area $A$
of the black hole satisfies $dA \ge 0$ (because the angular momentum 
parameter $a$ decreases, and consequently $r_+$ increases).
In the final stage of an idealized 
process, the mass of the black hole becomes the irreducible mass $M_{irr}$ 
\cite{CR}, defined by the relation $m^2=M_{irr}^2+J^2/(4M_{irr}^2)$, and
the Kerr black hole becomes a Schwarzschild black hole. The expression of the
irreducible mass is

\begin{equation}
M_{irr}={1\over 2} \sqrt{r_+^2+ a^2}\,.
\label{5-1-4}
\end{equation}

An analysis of various gravitational energy expressions for the Schwarzschild
and Kerr black holes has been made long time ago in Ref. \cite{Bergqvist}. 
Taking into account various expressions for the gravitational mass, it was 
concluded that the mass contained within the event horizon of the 
Schwarzschild black hole is $2m$, for all considered definitions.
The same value is obtained by the quasi-local expressions (as, for instance,
the Brown-York expression \cite{BY}).
Likewise, one expects that the mass contained within the external event 
horizon at the final stage of the complete Penrose
process, when the black hole becomes non-rotating, is given by $2M_{irr}$,
and the gravitational energy contained within the external event horizon of
the black hole is, therefore, $2M_{irr}c^2$. In similarity to the 
Schwarzschild space-time, the total gravitational energy of the Kerr 
space-time finally becomes $M_{irr}c^2$, after the complete Penrose
process.

We have already shown that the definition for the gravitational energy in the
TEGR yields an expression for the energy contained within the external event 
horizon of the Kerr black hole that is strikingly close to $2M_{irr}c^2$. In
this Section, we will review this result. 

We proceed to establish the frame that satisfies Schwinger's time gauge
condition. Here we assume $c=1$, but the notation will be that of subsection
2.4. As we mentioned above, the 
frame must be defined such that the radial coordinate $r$ runs from $r_+$
to infinity, i.e., the frame must be defined in the whole region outside the 
external event horizon, and consequently also inside the ergosphere of the 
black hole. The external surface of the ergosphere is defined by  
$r=r_+^*=m+ \sqrt{m^2-a^2\cos^2 \theta}$. Inside the 
ergosphere all observers are necessarily dragged in circular motion by the 
gravitational field. The four-velocity of observers that circulate around the
black hole, outside the external horizon, under the action of the 
gravitational field of the Kerr space-time, is given by 

\begin{equation}
u^\mu(t,r,\theta,\phi)
={{\rho \Sigma}\over{(\psi^2\Sigma^2+\chi^2\sin^2\theta)^{1/2}}}
(1,0,0,{\chi \over {\Sigma^2}})\,,
\label{5-1-5}
\end{equation}
where all functions are defined by Eqs. (\ref{1-5-1}) and (\ref{1-5-2}).
It is possible to show that if we restrict the radial coordinate to $r=r_+$,
the $\mu=3$ component of Eq. (\ref{5-1-5}) becomes

$${\chi \over {\Sigma^2}}= {a\over {2mr_+}} =
{a \over {a^2+r_+^2}}= \Omega_H\,,$$
The quantity

\begin{equation}
\omega(r)= -{{g_{03}}\over{g_{33}}}={{\chi}\over{ \Sigma^2}}\,,
\label{5-1-6}
\end{equation}
is the dragging angular velocity of inertial frames.

The tetrad fields {\bf (i)} that are adapted to observers whose 
four-velocities are given by Eq. (\ref{5-1-5}), i.e., for which 
$e_{(0)}\,^\mu =u^\mu$, and consequently defined in the region $r > r_+$,
{\bf (ii)} whose $e_{(i)}\,^\mu$ components in Cartesian coordinates are 
asymptotically oriented along the unit vectors $\hat{\bf x}$, $\hat{\bf y}$,
$\hat{\bf z}$, and {\bf (iii)} that is asymptotically flat, is given by

\begin{equation}
e_{a\mu}=\pmatrix{-A&0&0&0\cr
B\sin\theta\sin\phi
&C\sin\theta\cos\phi& D\cos\theta\cos\phi&-F\sin\theta\sin\phi\cr
-B\sin\theta\cos\phi
&C\sin\theta\sin\phi& D\cos\theta\sin\phi& F\sin\theta\cos\phi\cr
0&C\cos\theta&-D\sin\theta&0}\,,
\label{5-1-7}
\end{equation}
where

\begin{eqnarray}
A&=& {{(g_{03}g_{03}-g_{00}g_{33})^{1/2}}\over{(g_{33})^{1/2}}}\,,
\nonumber \\
B&=&-{{ g_{03}}\over {(g_{33})^{1/2} \sin\theta}}\,, \nonumber \\
C&=&(g_{11})^{1/2}\,, \nonumber \\
D&=&(g_{22})^{1/2}\,, \nonumber \\
F&=& {{(g_{33})^{1/2}}\over {\sin\theta}}\,.
\label{5-1-8}
\end{eqnarray}
These tetrad fields are the unique configuration that satisfies the above
conditions, since six conditions are imposed on $e^a\,_\mu$. It satisfies
Schwinger's time gauge condition $e_{(i)}\,^0=0$. This frame allows
observers to reach the vicinity of the external event horizon of the Kerr 
black hole. Therefore we may evaluate the gravitational energy 
contained within any surface $S$ determined by 
the condition $r > r_+$, and in particular in the limit $r \rightarrow r_+$.
Expression (\ref{5-1-7}) is precisely the same set of 
tetrad fields (Eq. (4.9)) considered in Ref. \cite{Maluf14}.

The energy contained within the external event horizon of the black hole is
calculated by means of the $a=(0)$ component of Eq. (\ref{11}) or (\ref{22}),

\begin{equation}
P^{(0)}=E=-\oint_S dS_i \Pi^{(0)i}=
-\oint_S d\theta d\phi\,\Pi^{(0)1}(r,\theta,\phi)\,.
\label{5-1-9}
\end{equation}
$S$ is a surface of constant radius determined by the condition $r=r_+$
(recall that we are assuming $c=1$). After a number of algebraic
calculations we obtain \cite{Maluf14}

\begin{equation}
E=m\biggl[ {\sqrt{2p}\over 4}+{{6p-q^2}\over {4q}}
\ln \biggl({{\sqrt{2p}+q}\over p}\biggl) \biggr]\,.
\label{5-1-10}
\end{equation}
The quantities $p$ and $\lambda$ are defined by

$$p=1+\sqrt{1-q^2}\,, \ \ \ \ \ \ a= q\, m\,, \ \ \ \ \ \ 
0\le q \le 1\,.$$
Note that when $q=0$, we have $E=2m$.
Equation (\ref{5-1-10}) is functionally very different from  
$2M_{irr}=\sqrt{r_+^2+ a^2}$. However, the two expressions are very similar,
as we can verify in Figure 1.

\begin{figure}[h]
\centering
\includegraphics[width=0.6\textwidth]{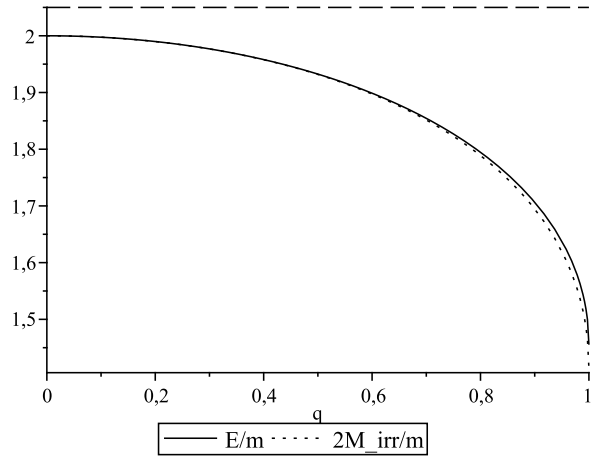}
\caption{Gravitational energy in the interior of the external event horizon of
the Kerr black hole, as function of $q=a/m$.}
\end{figure}

In Figure 1, we have plotted (i) $E/m$, where $E$ is given by Eq. 
(\ref{5-1-10}), and (ii) $2M_{irr}/m$, as functions of $a/m=q$. Both 
curves are parametrized by $q$, 
which varies from 0 to 1. The upper curve
represents Eq. (\ref{5-1-10}), and the lower curve represents $2M_{irr}$. The
almost coincidence between the two expressions is striking, and is perhaps 
the major achievement of definition (\ref{22}). It shows that 
Eq. (\ref{5-1-10}) is in very close agreement with $2M_{irr}$, as expected. 
This result supports the idea of localization of the gravitational 
energy.

It is not the purpose of this review to address or critique alternative
approaches that deal with the same investigation of this Section, but we may
just compare our result with the one 
obtained by other means, notably the quasi-local 
expression for the gravitational energy. A recent proposal for the 
quasi-local gravitational energy was given by Wang and Yau 
\cite{WY1,WY2}, and applied to the Kerr space-time \cite{WY3} to address 
precisely the irreducible mass of the black hole. In the context of the 
latter references, it was made an analysis similar to the one above that
led to Figure 1. However, due to mathematical reasons and difficulties, the
analysis can be carried out only in the range $0 \leq a \leq \sqrt{3}m/2$ for 
the values of $a=J/m$. A latter attempt to circumvent this problem was made
in Ref. \cite{Maciej}, by means of an even different approach to quasilocal
mass. By inspecting Figure 3 of the latter reference, which is the equivalent
to Figure 1 of the present analysis, we see a
sharp and unnatural discontinuity of the curve at the value $a=\sqrt{3} m/2$.
In addition, the value of the mass within the horizon exceeds $2m$ for values
of $a$ close to $m$. In our opinion, these are difficulties of the whole
method based on quasi-local expressions for the gravitational energy, which
cannot reproduce $2M_{irr}$ in the whole range of values of the parameter 
$a$. Furthermore, the latter approach to gravitational energy is based only 
on the metric tensor, and for this reason it should be independent of frames
or observers, but not so long ago the need to introduce ``quasi-local 
observers" was put forward in Ref. \cite{Wang}. This attempt shows that, one
way or another, observers and frames are necessary to address the notion of 
gravitational energy. After all, the observed mass of the black holes depends
on the frame, in case the black hole is moving with respect to the frame. An
implicit dependence on the frame already takes place at the very initial
considerations of the Schwarzschild or Kerr line elements: the parameter $m$
is the total mass of black hole in a frame where the black hole 
{\it is at rest}.

\section{Distribution of gravitational energy in the Schwarzschild-de Sitter space-time}

The main purpose of this section is to review the fact that a cosmological
constant, even in a simple and idealised model such as the Schwarzschild-de
Sitter space-time, induces an extremely high density of energy in the 
vicinity of the cosmological horizon, determined by a surface of constant 
radius $r\approx R$, where $R=\sqrt{3/\Lambda}$, and $\Lambda$ is a positive
cosmological constant. In the pure de Sitter space-time, the cosmological
horizon is determined exactly by $r=R$. To our knowledge, the result of this
section can only be obtained in the realm of the TEGR.

The line element of the Schwarzschild-de Sitter space-time is given by

\begin{equation}
ds^2=-\alpha^2\,dt^2+{1\over \alpha^2}dr^2 + r^2d\theta^2+
r^2\sin^2\theta d\phi^2\,,
\label{27}
\end{equation}
where

\begin{equation}
\alpha^2=1-{{2m}\over r} -{{r^2}\over {R^2}}\,.
\label{28}
\end{equation}
Here we are considering natural units and making 
the speed of light $c=1$. The Schwarzschild-de Sitter
space-time has been considered in the TEGR in Ref. \cite{Maluf16}. In this
Section we make use of the notation and some results of Ref. \cite{Maluf11}.
We will restrict the considerations to the region between the Schwarzschild
and cosmological horizons. The set of tetrad fields adapted to stationary 
observers in the space-time is given by

\begin{equation}
e_{a\mu}=\pmatrix{-\alpha&0&0&0\cr
0&\alpha^{-1}\sin\theta\,\cos\phi&r\cos\theta\,\cos\phi
&-r\sin\theta\,\sin\phi\cr
0&\alpha^{-1}\sin\theta\,\sin\phi&r\cos\theta\,\sin\phi
&r\sin\theta\,\cos\phi\cr
0&\alpha^{-1}\cos\theta&-r\sin\theta&0\cr}\,.
\label{29}
\end{equation}

It is not difficult to calculate the
radial  (inertial, non-gravitational) accelerations that are necessary to
maintain the frame static in space-time. They are given by

\begin{eqnarray}
\phi_{(0)(1)}&=&{{d\alpha}\over{dr}}\sin\theta\cos\phi\,,\nonumber \\
\phi_{(0)(2)}&=&{{d\alpha}\over{dr}}\sin\theta\sin\phi\,,\nonumber \\
\phi_{(0)(3)}&=&{{d\alpha}\over{dr}}\cos\theta\,.
\label{34}
\end{eqnarray}
We define the inertial acceleration vector $\Phi$ as 

\begin{equation}
\Phi(r)=(\phi_{(0)(1)}, \phi_{(0)(2)},\phi_{(0)(3)})\equiv\phi(r) \hat{r}
={{d\alpha}\over{dr}} \hat{r} \,,
\label{35}
\end{equation}
where $\hat{r}=(\sin\theta\cos\phi, \sin\theta\sin\phi, \cos\theta)$, and 

\begin{equation}
\label{36}
{{d\alpha}\over {dr}}={1\over \alpha}\biggl({m\over r^2} - 
{r\over R^2}\biggr)\,.
\end{equation}
Close to the Schwarzschild horizon we have $d\alpha/dr >0$, and close to the
cosmological horizon, $d\alpha /dr <0$, as it should be to maintain the 
frame static in space-time \cite{Maluf11} (close to the Schwarzschild horizon
the gravitational acceleration is attractive, and close to the de Sitter
horizon it is repulsive).

In view of the spherical symmetry of the physical configuration, the 
gravitational energy obtained from Eq. (\ref{22}) is reduced to

\begin{equation}
\label{37}
P^{(0)}=-\oint_S dS_1\,\Pi^{(0)1}\,,
\end{equation}
where $dS_1=d\theta d\phi$.
After a number of simple calculations, we find that 

\begin{equation}
\label{38}
-\Pi^{(0)1}=4ke\,\Sigma^{(0)01}={1\over {4\pi}}r\,\sin\theta
\biggl( 1-\sqrt{ 1-{{2m}\over {r_0}}-{{r_0^2}\over {R^2}}} \biggr)\,,
\end{equation}
and thus the energy contained
within a surface of constant radius $r_0$ is given by

\begin{equation}
\label{39}
P^{(0)}=r_0\biggl( 1-\sqrt{ 1-{{2m}\over {r_0}}-{{r_0^2}\over {R^2}}}
\biggr)\,.
\end{equation}

In the following, we will simplify the considerations and make $m=0$. 
By also making $r_0=R$ in Eq. (\ref{39}), we find \cite{Maluf16}

\begin{equation}
\label{40}
P^{(0)}=R=\sqrt{3\over \Lambda}\,,
\end{equation}
which is the total gravitational energy contained in the interior of the 
cosmological horizon. Returning to Eq. (\ref{37}), we note that it can be
written as a volume integral,

\begin{equation}
\label{41}
P^{(0)}=-\int_V dr\,d\theta d\phi \partial_r \Pi^{(0)1}\,,
\end{equation}
where $V$ is any spherical volume of radius $r_0$ such that
$r_0\leq R$. Now we take the radial
derivative of $\Pi^{(0)1}$ in the equation above, and integrate in the 
angular variables. The resulting expression is

\begin{equation}
\label{42}
P^{(0)}=\int_V dr\,\varepsilon (r)\,,
\end{equation}
where $\varepsilon(r)$ expresses the gravitational energy density in the pure
de Sitter space, i.e., $\varepsilon(r)\,dr$ is the gravitational energy 
contained within the shells of radii $r$ and $r+dr$. It reads \cite{Maluf16}

\begin{equation}
\label{43}
\varepsilon(r)=1\,+\,
{{2\beta^2-1}\over \sqrt{1-\beta^2}}\;,
\end{equation}
where $\beta^2=r^2/R^2$.
Note that if $\beta \rightarrow0$, then $\varepsilon(0)\rightarrow0$ and if 
$\beta\rightarrow 1 $, $\varepsilon$ diverges. These features are shown
in Figure 2.

\begin{figure}[h]
\centering
\includegraphics[width=0.6\textwidth]{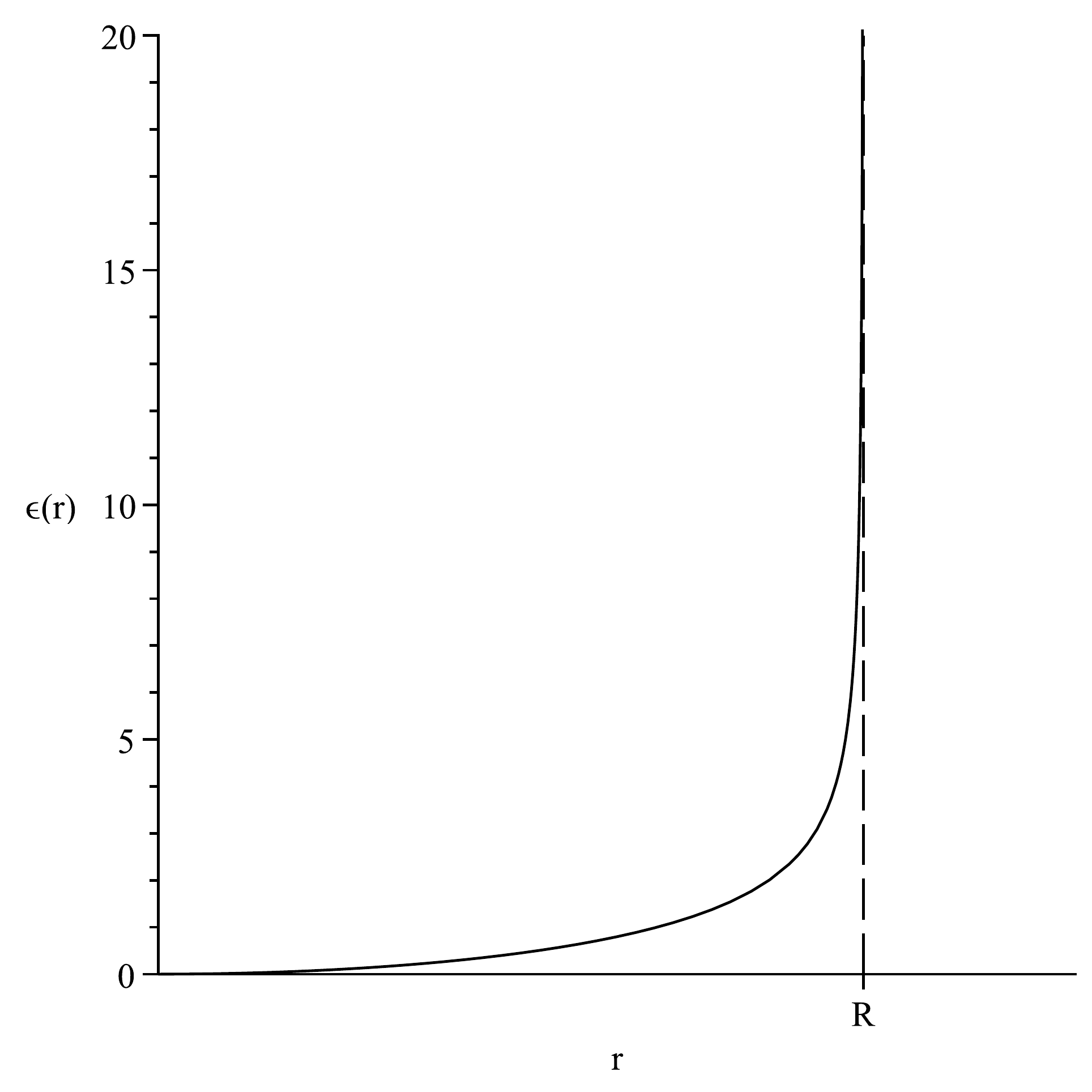}
\caption{The gravitational energy density in the de Sitter space-time as
a function of the radial coordinate only, after integration on the angular
variables.}
\end{figure}

The total energy contained inside the surfaces of 
radii $0.1R\;,\;0.5R\;,\;0.9R$ are given by
$E_g=5.01\times 10^{-4}R\;,\;E_g=0,067R\;,\;E_g=0.51R$,
respectively. Thus, almost half of the gravitational 
energy is located between $\beta=0.9$ and $\beta=1$.

We see that the cosmological constant induces a concentration of 
gravitational energy in the vicinity of the cosmological horizon. Therefore,
the existence of an extremely high density of energy located very far from
idealised observers, and which could play the role of the so called dark 
energy in cosmological models, is explained by the definition of 
gravitational energy in the TEGR, even in the absence of matter fields.

It is also very interesting to observe that the cosmological constant gives
rise to a classical vacuum energy that is not uniform over the 3-dimensional
space, contrary to what is assumed in ordinary quantum field theories in
flat space-times. The presence of a very tiny cosmological
constant may not be perceived on neighbourhoods such as our planetary
system, but its existence modifies the vacuum energy density on a 
cosmological scale. By inserting the physical constants $c$ and $G$ (the
gravitational constant) into Eq. (\ref{40}), we have 

\begin{equation}
\label{44}
P^{(0)}={E\over c}= {c^3\over G}\sqrt{3\over \Lambda}\,,
\end{equation}
where $\Lambda\approx 10^{-56}\,cm^{-2}$ and $G=6,67\times 10^{-8}
{cm^3\over {g\cdot s^2}}$. It follows that

\begin{equation}
\label{45}
E=E_{\small{vacuum}}\approx21\times 10^{76}\,{{g \cdot cm^2}\over s^2}\,.
\end{equation}
This is a significant value that should modify some assumptions in ordinary
quantum field theories, regarding the uniformity in space of the vacuum 
energy density, in case the space-time is endowed with a positive 
cosmological constant.

\section{Gravitational energy-momentum in the Bondi-Sachs space-time}

The space-time of a radiating gravitational field configuration is described
by the Bondi-Sachs line element. The latter is usually presented in the 
($u$, $r$, $\theta$, $\phi$) coordinates, where $r$ is the ordinary radial 
variable and $u$ is the retarded time, $u=t-r$ ($r$, $\theta$ and $\phi$
are spherical coordinates). The main quantities in the 
Bondi-Sachs line element are the functions $M(u,\theta,\phi)$, which is the
{\it mass aspect}, and $c(u,\theta,\phi)$, $d(u,\theta,\phi)$, which are 
related to the {\it news functions}. 

The Bondi-Sachs line element is primarily a model for
isolated astrophysical configurations that loose energy in the form of 
gravitational radiation. The space-time 
around these configurations is not strictly asymptotically flat because 
although the metric tensor components fall off in the expected way as 
$g_{\mu\nu}=\eta_{\mu\nu}+h_{\mu\nu}(1/r)$ in the asymptotic limit
$r\rightarrow \infty$, the time derivative of $h_{\mu\nu}$ is of order 
$1/r$ at spacelike infinity in Cartesian coordinates. 
The first significant investigation of this physical system
was made by Bondi and collaborators \cite{Bondi}, who established the
coordinates and notation that are currently employed in the analysis of
gravitational radiation. 
The line element obtained by Bondi and collaborators
is constructed out of the functions $M(u,\theta)$ and $c(u,\theta)$ only.
These functions do not depend on the azimuthal coordinate 
$\phi$ because the line element is axially symmetric. The time derivative 
$\partial_0 c=\partial c/\partial u$ is identified as the first news 
function.
Bondi's line element was later generalised by Sachs \cite{Sachs}, who
abandoned the axial symmetry and obtained the most general metric tensor that 
describes gravitational radiation at spacelike and null infinities. In the 
work by Sachs, there also appears the function $d(u,\theta,\phi)$, that 
yields the second news function $\partial_0 d$, and all functions $(M,c,d)$
now depend also on the angular variable $\phi$.

The two news functions are interpreted as the radiating degrees of freedom of 
the gravitational field, and this single feature justifies the importance of
the metric tensor that describes the Bondi-Sachs space-time.
The mathematical expression of the Bondi-Sachs metric tensor is not 
given in closed mathematical form. It is presented in powers of the radial 
coordinate $r$. There are very good review articles that clarify the several 
aspects of the subject, and also explain the emergence of the Bondi-Sachs 
energy-momentum vector \cite{Trautman,Pirani,Sachs-2,Goldberg}.

The Bondi-Sachs energy-momentum $m_\mu(u)$ 
is constructed out of the mass aspect
$M(u,\theta,\phi)$ only, i.e., it does not depend on the functions 
$c(u,\theta,\phi)$ and $d(u,\theta,\phi)$ (see, for instance, Eq. (4.4) of
Ref. \cite{Goldberg}). However, the news functions are expected to yield
real, observable manifestations of the gravitational field at spacelike or
null infinities, and so they should contribute to surface integrals, both at
spacelike and null infinities. For this reason, it is intriguing that the 
total Bondi-Sachs energy-momentum does not depend on the news function.

The ADM expression \cite{ADM} of the total gravitational energy-momentum 
is evaluated at spacelike infinity. However, the authors of Refs. 
\cite{Z1,Z2,Z3} have expressed the ADM energy-momentum in terms
of the Bondi-Sachs energy-momentum $m_\mu(u)$, which is linear in the 
integration of the mass aspect $M(u,\theta,\phi)$ at null infinity. 
More specifically, the authors of the latter
references rewrote the metric tensor of the Bondi-Sachs line 
element in the ordinary $(t,r,\theta,\phi)$ coordinates, and parametrized
the spacelike hypersurfaces by means of the standard time coordinate $t$.
As these authors explain, the prior
assumption is that the ADM energy-momentum is the past limit of the 
Bondi-Sachs energy-momentum, i.e.,  $m_\mu (- \infty)$.
They found, however, that the resulting expression for
the total ADM energy-momentum depends on the integration of the mass aspect
$M(u)$ when $u\rightarrow -\infty$, as expected (as well as on the spatial
orientations for the 3-momenta), 
but also depends on integrations of functions that depend on $c$ and 
$d$. Although the total ADM energy-momentum is 
strictly constructed for asymptotically flat space-times, the new and
modified expression for the Bondi-Sachs energy-momentum,
and its dependence on the functions $c$ and $d$,  is a very interesting and
reasonable result.

In this Section we will show that gravitational energy-momentum (\ref{11})
yields a quite interesting result. The latter expression renders the usual,
expected integration 
of the mass aspect, plus a term that is interpreted as the energy
of the gravitational radiation, since it depends only on the functions $c$
and $d$. This is a result that can only be achieved in the TEGR.
We will present the essential functions of the Bondi-Sachs 
space-time and the set of tetrad fields, but we will omit several long 
calculations that have already been presented in the literature 
\cite{Maluf17}. We will emphasize the results only.

The Bondi-Sachs line element is given by

\begin{eqnarray}
ds^2&=&g_{00}\,du^2+g_{22}\,d\theta^2+g_{33}\,d\phi^2 \nonumber \\
&{}&+2g_{01}\,du\,dr+2g_{02}\,du\,d\theta+2g_{03}\,du\,d\phi+
2g_{23}d\theta\,d\phi\,, 
\label{46}
\end{eqnarray}
where

\begin{eqnarray}
g_{00}&=&{V\over r}e^{2\beta}-r^2(e^{2\gamma}U^2\cosh 2\delta 
+e^{-2\gamma}W^2\cosh 2\delta +2UW \sinh 2\delta)\,, \nonumber \\
g_{01}&=& -e^{2\beta}\,, \nonumber \\
g_{02}&=&-r^2(e^{2\gamma}U\cosh 2\delta+W\sinh 2\delta)\,, \nonumber \\
g_{03}&=&-r^2\sin\theta(e^{-2\gamma}W\cosh 2\delta 
+U \sinh 2\delta)\,, \nonumber \\
g_{22}&=&r^2e^{2\gamma}\cosh 2\delta\,, \nonumber \\
g_{33}&=&r^2e^{-2\gamma}\cosh 2\delta\,\sin^2 \theta\,, \nonumber \\
g_{23}&=&r^2\sinh 2\delta\,\sin\theta\,.
\label{47}
\end{eqnarray}
We adopt the usual convention $(u,r,\theta,\phi)=(x^0,x^1,x^2,x^3)$.
The functions $\beta$, $\gamma$, $\delta$, $U$ and $W$ in the equations above
are not exact. They are given only in asymptotic form, in powers of $1/r$. 
We will dispense with the powers of $1/r$ of 
the field quantities that do not contribute to the calculations. Thus, the 
asymptotic form of the functions above are

\begin{eqnarray}
V&\simeq& -r+2M\,, \nonumber \\
\beta &\simeq& -{{c^2+d^2}\over {4r^2}}\,,\nonumber \\
\gamma &\simeq& {c\over r}\,, \nonumber \\
\delta &\simeq& {d\over r}\,, \nonumber \\
U&\simeq& -{{l(u,\theta,\phi)}\over r^2}\,, \nonumber \\
W&\simeq& -{{\bar{l}(u,\theta,\phi)} \over r^2}\,,
\label{48}
\end{eqnarray}
where 
$$l=\partial_2 c+2c\,\cot\theta+\partial_3 d\,\csc \theta\,,$$
$$\bar{l}=\partial_2 d +2d\,\cot\theta -\partial_3 c\csc \theta\,.$$
In the limit $r\rightarrow \infty$, the asymptotic form of the functions above
yield

\begin{eqnarray}
g_{00}&\simeq & -1+{{2M}\over r}\,, \nonumber \\
g_{01}&\simeq & -1+{{c^2+d^2}\over {2r^2}}\,,\nonumber \\
g_{02}&\simeq & l+{1\over r}(2cl +2d\bar{l} -p)\,, \nonumber \\
g_{03}&\simeq & \bar{l}\sin\theta + 
{1\over r}(-2c\bar{l}+2dl -\bar{p})\sin\theta\,, \nonumber \\
g_{22}&\simeq & r^2+ 2cr+2(c^2+d^2)\,, \nonumber \\
g_{33}&\simeq & \lbrack r^2- 2cr+2(c^2+d^2)\rbrack\sin^2\theta\,,\nonumber \\
g_{23}&\simeq & 2dr\sin\theta+{{4d^3}\over {3r}} \sin\theta\,.
\label{49}
\end{eqnarray}
The functions $p$ and $\bar{p}$ are defined in Refs. \cite{Z1,Z2}. They 
depend on functions that are not defined above. However, they will not 
contribute to the final expressions, and for this reason we will not present 
their definitions here.

The expressions of the contravariant components of the metric tensor are
calculated by means of the usual procedure out of Eqs. (\ref{47}) (and 
not out of Eqs. (\ref{49})). The inverse components are given by 
$g^{\mu\nu}=(-1)^{\mu+\nu}(1/g)\,M_{\mu\nu}$,  where 
$g=-g_{01}^2(g_{22}g_{33}-g_{23}^2)$ is the determinant of the metric tensor,
and $M_{\mu\nu}$ is the co-factor of the $\mu\nu$ component (we have taken 
into account all necessary powers of $1/r$ of the functions given in 
Eq. (\ref{48})).
We find

\begin{eqnarray}
g^{00}&=& g^{02}=g^{03}=0\,,\nonumber \\
g^{01}&\simeq & -1-{{c^2+d^2}\over {2r^2}}\,,\nonumber \\
g^{11}&\simeq &1-{{2M}\over r}\,,\nonumber \\
g^{12}&\simeq &{l\over r^2}\,,\nonumber \\
g^{13}&\simeq &{{\bar{l}\sin\theta}\over r^2}\,,\nonumber \\
g^{22}&\simeq &{1\over r^2}\,,\nonumber \\
g^{33}&\simeq &{1\over {r^2\sin^2\theta}}\,,\nonumber \\
g^{23}&\simeq & -{{2d}\over {r^3\sin^2\theta}}\,.
\label{50}
\end{eqnarray}

In order to investigate the gravitational radiation in the context of the 
Bondi-Sachs space-time, it is convenient to choose a set of tetrad
fields adapted to static observers in space-time. The worldline of these
observers is characterized by the field of 4-velocities $u^\mu$ of the
observers such that $u^\mu=(u^0,0,0,0)$. Following the procedure discussed
in Section 2, we identify the $a=(0)$ component
of the frame $e_ a\,^\mu$ as $e_{(0)}\,^\mu=u^\mu$.
The three other components of the frame, $e_{(i)}\,^\mu$, are orthogonal
to $e_{(0)}\,^\mu$, and may be oriented in the three-dimensional space 
according to the symmetry of the physical system, in case there are 
privileged directions in the 3-space.
It follows from the requirement $e_{(0)}\,^\mu=u^\mu$ that a static observer
in space-time must satisfy the
conditions $e_{(0)}\,^i(t,x^k)=(0,0,0)$. It is easy to verify, by means of a
coordinate transformation, that in terms of the retarded time $u$ we also 
have $e_{(0)}\,^i(u,x^k)=(0,0,0)$. 

The Bondi-Sachs space-time is not axially
symmetric. It is clear from the mathematical structure of the line element
that there are no distinguished directions at spacelike
infinity. Since we will evaluate surface integrals at spacelike infinity, 
i.e., we will be interested only in total quantities (we will integrate
over a surface $S$ determined by $r=$ constant, for $r$ finite but 
sufficiently distant from the source),
any set of tetrad fields that satisfy the asymptotic expansion 
$e_{a\mu}\simeq \eta_{a\mu}+(1/2)h_{a\mu}(1/r)$
in Cartesian coordinates when $r\rightarrow \infty$, and that satisfy the 
conditions $e_{(0)}\,^i(u,r,\theta,\phi)=0$, will serve our purposes.
For such a frame, $e_{(1)}\,^\mu$, $e_{(2)}\,^\mu$ and $e_{(3)}\,^\mu$ will 
define the usual unit frame vectors in the $x$, $y$ and $z$ directions, 
respectively, in the limit $r\rightarrow \infty$, provided $e_a\,^\mu$ is 
constructed in Cartesian coordinates. These conditions establish a 
consistent reference frame at spacelike infinity.

In spite of the intricacies of the Bondi-Sachs line element, it is
possible to establish a simple set of tetrad fields that 
yields Eqs. (\ref{46}) and (\ref{47}), and that satisfy the conditions 
$e_{(0)}\,^i=0$. Note that $e_{(0)}\,^i=0$ implies $e^{(i)}\,_0=0$. One
set of tetrad fields that satisfy these requirements, and has the asymptotic
form $e_{a\mu}\simeq\eta_{a\mu}+(1/2)h_{a\mu}$ at spacelike infinity, is 
given in $(u,r,\theta,\phi)$ coordinates by \cite{Maluf17} 

\begin{eqnarray}
e_{(0)\mu}&=&(-A,-E,-F,-G)\,, \nonumber \\
e_{(1)\mu}&=& (0, \,
B_1\sin\theta \cos\phi+B_2\cos\theta \cos\phi-B_3\sin\theta \sin\phi\,, 
\nonumber \\
& & C_1 r \cos\theta \cos\phi-C_2 \sin\theta \sin\phi\,, \nonumber \\
& & -Dr\sin\theta \sin\phi)\,,\nonumber \\
e_{(2)\mu}&=& (0, \,
B_1\sin\theta \sin\phi+B_2\cos\theta \sin\phi+B_3\sin\theta \cos\phi\,, 
\nonumber \\
& & C_1 r \cos\theta \sin\phi-C_2 \sin\theta \cos\phi\,, \nonumber \\
& & Dr\sin\theta \cos\phi)\,,\nonumber \\
e_{(3)\mu}&=&(0,\,B_1\cos\theta-B_2\sin\theta, -C_1r\cos\theta,0)\,.
\label{51}
\end{eqnarray}

The quantities $A,B_1,B_2,B_3,C_1,C_2,D,E,F,G$ are determined by requiring
that $e_{a\mu}$ yields the metric tensor components (\ref{49}) according to
$e_{a\mu}e_{b\nu}\eta^{ab}=g_{\mu\nu}$. The determination of
exact form of these quantities in terms of the metric tensor components 
(\ref{47}) is not so simple. These quantities must satisfy the following
equations,

\begin{eqnarray}
-A^2&=& g_{00}\nonumber \\
-AE&=&g_{01} \nonumber \\
-AF&=&g_{02} \nonumber \\
-AG&=&g_{03} \nonumber \\
-E^2+B_1^2+B_2^2+B_3^2\sin^2\theta&=&g_{11}=0\nonumber \\
-EF+B_2(C_1r)+B_3C_2\sin^2\theta&=&g_{12}=0\nonumber \\
-EG+B_3(Dr)\sin^2\theta&=&g_{13}=0 \nonumber \\
-F^2+(C_1r)^2+C_2^2\sin^2\theta&=&g_{22} \nonumber \\
-G^2+(Dr)^2\sin^2\theta&=&g_{33} \nonumber \\
-FG+C_2(Dr)\sin^2\theta&=& g_{23}\,.
\label{52}
\end{eqnarray}

In view of the fact that we need the components of the torsion tensor only 
in the asymptotic limit $r \rightarrow \infty$, we restrict the 
considerations to quantities whose power of $1/r$ are actually needed in the
calculations. Thus, we consider terms of power up to $(1/r)^2$ only. 
Quantities of order $(1/r)^n$, with $n\geq 3$, do not contribute to the 
final, total expressions, and will be neglected. We find

\begin{eqnarray}
A&\simeq& 1-{M\over r}\,, \nonumber \\
E&\simeq& 1+{M\over r}\,,\nonumber \\ 
F&\simeq& -l-{1\over r}(2cl+2d\bar{l}+Ml-p)\,, \nonumber \\
G&\simeq& -\sin\theta \biggl[ \bar{l}+
{1\over r}(-2c\bar{l}+2dl+M\bar{l}-\bar{p})\biggr]\,, \nonumber \\
B_1 &\simeq& 1+{M\over r}\,, \nonumber \\
B_2 &\simeq& -{l\over r}-{1\over r^2}(2Ml+cl-p)\,, \nonumber \\
B_3&\simeq& -{1\over {\sin\theta}}\biggl[ {{\bar{l}\over r}}+{1\over r^2}
(2M\bar{l}-c\bar{l}+2dl-\bar{p})\biggr]\,,\nonumber \\
C_1&\simeq& 1+{c\over r}+{1\over r^2}
\biggl[{{l^2}\over 2}+c^2-d^2\biggr]\,,\nonumber \\
C_2&\simeq& {1\over {\sin\theta}}\biggl[ 2d+{1\over r}(l\bar{l}+2cd)\biggr]\,.
\nonumber \\
D&\simeq& 1-{c\over r}+{1\over r^2}\biggl( {{\bar{l}^2\over 2}}+
c^2+d^2\biggr)\,.
\label{53}
\end{eqnarray}

The expressions above completely fix the set of tetrad fields given by 
Eq. (\ref{51}). We note that the $\sin\theta$ in the denominator of
$B_3$ and $C_2$ in the expressions above does not lead to a divergence
of $e_{a\mu}$ on the $z$ axis of the coordinate system, when $\theta =0$. In 
Eq. (\ref{51}) these quantities are multiplied by $\sin\theta$. It can be 
shown that both in $(u,x,y,z)$ or in $(t,x,y,z)$ coordinates, the set 
of tetrad fields 
given by (\ref{51}) is everywhere smooth in the three-dimensional space
(except possibly at $r=0$, but the functions (\ref{48}) are not exactly 
defined in the neighbourhood of $r=0$).

As we mentioned earlier, we will skip many long calculations, and present 
the main results only. Some important steps in the calculations are presented
in Ref. \cite{Maluf17}. After many calculations and simplifications, we 
obtain the total gravitational energy by means of Eq. (\ref{11}). It reads

\begin{equation}
P^{(0)}=4k\int_0^{2\pi} d\phi \int_0^{\pi}d\theta \sin\theta
\biggl[M+\partial_0 F\biggr]\,, 
\label{54}
\end{equation}
where

\begin{equation}
F=-{1\over 4}\biggl(l^2 + \bar{l}^2 \biggr) +{1\over 2}c^2 +d^2\,.
\label{55}
\end{equation}
The quantities under integration in the expression above are the only ones
that contribute to the surface integral in the limit $r\rightarrow \infty$.

The new term $\partial_0 F$ generalises the standard Bondi-Sachs energy.
As expected, the latter depends not only on the integral of the mass aspect
$M(u,\theta,\phi)$, but also on the functions $c(u,\theta,\phi)$ and 
$d(u,\theta,\phi)$.

The surface $S$ of integration in Eq. (\ref{11}) is usually a spacelike
surface of constant radius $r$. In the context of Eq. (\ref{54}), 
for finite values of the ordinary time $t$, the limit $r\rightarrow \infty$ 
corresponds to $u\rightarrow -\infty$, and therefore in Eq. (\ref{54}) 
we have $P^{(0)}(u\rightarrow-\infty)$. However, this is not a mandatory
condition.
We may consider very large but finite values of the radial coordinate
$r$ ($r\gg M$ and $r\gg \partial_0F$), as explained in Ref. \cite{Maluf17},
such that $M$ and $\partial_0 F$ remain the only contributions to the 
surface integral in Eq. (\ref{54}).
In this case, $P^{(0)}$ depends on arbitrary and finite values of $u$. The
fact is that real observers are never strictly located at spacelike infinity.

Equation (\ref{54}) is interpreted as the gravitational energy that
a static observer measures at very large distances from the source determined
by $M(u,\theta,\phi)$. Assuming the speed of light $c=1$ as well as $G=1$, 
the quantity 

\begin{equation}
E_{rad}={1\over {4\pi}}\int_0^{2\pi} d\phi \int_0^{\pi}d\theta \sin\theta\,
(\partial_0 F)\,,
\label{56}
\end{equation}
is interpreted as the energy of the gravitational radiation, i.e., a form
of energy detached from the source.

In the study of electrodynamics of moving but isolated sources, one finds that
the total electric field is composed of a term that is usually called the 
{\it near field}, which is determined by the charges of the source, and by
a term called {\it far field}, which is determined not only by the 
charge, but by the velocity and acceleration of the charge. This latter
component of the electric field falls off as $1/r$ at spacelike infinity, 
and represents the first order contribution of the electric field at large
distances (it contributes to the Poynting vector, together with the 
corresponding magnetic field component). 
By making an analogy with expressions (\ref{55}) and (\ref{56}),
we may say that $E_{rad}$ given by (\ref{56}) is a form of 
{\it far energy}, that eventually may be detected and measured with the
increasing sophistication of the precision instruments presently used in the
large detectors of gravitational waves.

The gravitational momenta is calculated in a way very similar to the 
gravitational energy. Again, the algebraic calculations are somewhat long,
but pose no special mathematical difficulty. The gravitational momenta
$P^{(i)}\;$ $(i=1,2,3)$ 
may be presented in a simplified form by first defining the 
standard textbook vectors, 

\begin{eqnarray}
\hat{r}^i&=&(\sin\theta\cos\phi,\, \sin\theta\sin\phi,\,\cos\theta)\,, 
\nonumber \\
\hat{\theta}^i&=&(\cos\theta\cos\phi,\, \cos\theta\sin\phi,\, -\sin\theta)\,, 
\nonumber \\
\hat{\phi}^i&=&(-\sin\phi,\, \cos\phi,\,0)\,.
\label{57}
\end{eqnarray}
The final expressions are given by

\begin{eqnarray}
P^{(i)}&=& 4k\int_0^{2\pi}d\phi \int_0^{\pi}d\theta \sin\theta \biggl[
(M+\partial_0 F) \hat{r}^i \nonumber \\
&& +{1\over 4} (l\partial_0 M)\hat{\theta}^i +
{1\over 4}(\bar{l}\partial_0 M)\hat{\phi}^i \biggr]\,.
\label{58}
\end{eqnarray}

Altogether, $P^{(0)}$ and $P^{(i)}$ constitute the Bondi-Sachs 
energy-momentum in the realm of the TEGR.
One important observation is the following. In view of the field equations
for the metric tensor (\ref{46}), it is known that
the time derivative $\partial_0 M$ may be written
in terms of the time derivatives of the functions $c$ and $d$ 
\cite{Sachs,Sachs-2} according to

\begin{equation}
\partial_0 M=-\lbrack (\partial_0c)^2+(\partial_0 d)^2\rbrack
+{1\over 2}\partial_0\biggl(\partial_2 l+l\cot\theta+
{{\partial_3 \bar{l}}\over {\sin\theta}}\biggr)\,.
\label{59}
\end{equation}
If we evaluate all integrals in $P^a$ in the limit $r\rightarrow \infty$,
we are actually taking the limit $u\rightarrow -\infty$, as we discussed
above. If, in addition, we assume that the
news functions satisfy the initial conditions

\begin{equation}
\partial_0 c(-\infty)=0\,, \;\;\;\;\;\; \partial_0d(-\infty)=0\,,
\label{60}
\end{equation}
then the total energy-momentum $P^a=(P^{(0)},P^{(i)})$ given by (\ref{54}) and
(\ref{58}) reduces to the well known expression for the Bondi-Sachs 
energy-momentum given in Ref. \cite{Goldberg}, for instance.
These initial conditions are physically reasonable. They are related to
the fact that the physical configuration is not initially radiating (i.e.,
the radiation starts at a finite instant of the retarded time $u$), but
they are not mandatory. We asserted above that we can 
evaluate $(P^{(0)},P^{(i)})$ over a surface $S$ of integration sufficiently 
far from the source (so that only terms of order $1/r$ and $1/r^2$ contribute
to the calculations), such that the surface $S$ is large but 
\underline{finite}. This 
fact allows us to dispense with the initial conditions given by 
Eq. (\ref{60}), since the retarded time $u$ is not strictly $-\infty$, and 
treat Eq. (\ref{56}) as, indeed, the energy of the gravitational radiation.

As a final remark, we mention that in Ref. \cite{Maluf17} we have simplified
the analysis to an axially symmetric space-time, constructed an analytic
expression for the news function $c(u,\theta)$ and obtained an expression for
the energy of gravitational radiation $E_{rad}$ as a somewhat complicated 
infinite sum. We believe that this issue could be re-analysed, since the 
resulting expression for $E_{rad}$ is a sum of finite values, and
maybe it could be simplified to an infinite sum where each term can
be easily interpreted.

\section{Plane gravitational waves, the kinetic energy of free particles and localization of the gravitational energy}

As we asserted at the Introduction of this review, in a series of 
publications \cite{Maluf4,Maluf5,Maluf6,Maluf7} we have shown that the
action of plane-fronted gravitational waves (pp-waves) on free particles 
may increase or decrease the kinetic energy of the particle. The literature
on pp-waves is really vast, as we have indicated in the above references.
It is important to recall that pp-waves are very simple 
exact solutions of the
vacuum Einstein's field equations, and also of the teleparallel field 
equations (\ref{5-1}) and (\ref{5}). A very good analysis on this issue is 
presented in Refs. \cite{Ehlers,Ehlers-2}. As for any simple solutions of
field equations or of equations of motion in physics, these waves are 
idealizations of more general realistic configurations, but no doubt, the 
latter is expected to 
display the basic features of the idealized physical configurations. 
The belief is that pp-waves may exist in nature not exactly described by the
simple mathematical functions that we will present ahead, but by a more 
elaborate and realistic mathematical description.

We recall that free particles or free material systems are here 
understood as physical systems that are subject to the gravitational field
only, and not to any external inertial forces. 
The consequence of the action of pp-waves on free particles is two fold, at 
least. First, since free particles gain or loose energy with the passage of
the wave, so does the wave loose or gain energy from the 
particle. This feature
allows the wave to travel in space for periods of time such as billions of
years, without dissipating. The waves that are currently being observed in
the large terrestrial detectors are, supposedly, waves that have travelled
for such long periods of time. 

The second consequence is that since the free particles are idealized 
pointwise objects, the energy transfer is localized in space. This fact
demonstrates that the gravitational energy is localized. As we asserted at
the Introduction, it does not make sense to relate this energy transfer to
any spacelike 2-surface of arbitrary radius that envelops the particle.
The energy transfer is effectively pointwise, and this is the simplest 
description of the physical process. 

In the following, we will recall the exposition presented in Ref. 
\cite{Maluf6}, in the study of the memory effect due to the pp-waves. The 
space-time before and after the
passage of the wave is assumed to be flat or nearly flat, for suitable
amplitudes of the waves. The metric tensor that describes the plane wave
space-time has a well known and simple structure. We assume that
wave propagates along the $z$ direction. Then, the line element reads 

\begin{equation}
ds^2=dx^2+dy^2+2du\,dv+H(x,y,u)du^2\,,
\label{61}
\end{equation}
in $(u,v,x,y)$ coordinates. In the flat space-time region, we may identify

\begin{equation}
u = {1\over {\sqrt{2}}}(z-t)\,,
\label{62}
\end{equation}

\begin{equation}
v = {1\over {\sqrt{2}}}(z+t)\,.
\label{63}
\end{equation}
The coordinates $(x,y)$ parametrize the planes parallel to the wave front. 
The line element depends essentially on a function $H(x,y,u)$. Here, the
retarded time is not given by $u$, but by $-u$.

The class of exact solutions considered here belong to the well known
Kundt family of space-times. They were studied in great detail by Ehlers and
Kundt \cite{Ehlers,Ehlers-2}.
The dependence of the function $H$ on the 
variable $u$ is arbitrary, which is a typical feature of 
solutions of wave equations. In view of this arbitrariness, the function $H$
may be chosen so that it describes a short burst of gravitational wave,
in which case the $u$ dependence of $H$ may be given by smooth Gaussians or 
derivatives of Gaussians. The function $H$ must only satisfy \cite{Ehlers} 

\begin{equation}
\nabla^2 H=\biggl( {{\partial^2} \over{\partial x^2}}+ 
{{\partial^2} \over{\partial y^2}} \biggr)H=0\,.
\label{64}
\end{equation}
We assume that far from an astrophysical source, a realistic 
gravitational wave may be approximated by 
an exact plane wave, exactly as we do in the study of
electromagnetic waves. 
The form of the line element (\ref{61}), in the 
coordinates $(u,v,x,y)$, was first presented by  Brinkmann \cite{Brinkmann}.

In order to study the geodesic equations, we need first rewrite the line
element above in $(t,x,y,z)$ coordinates. It reads

\begin{equation}
ds^2=\biggl({H\over 2} -1\biggr)dt^2+dx^2+dy^2+
\biggl({H\over 2}+1\biggr) dz^2-H\,dt dz\,.
\label{65}
\end{equation}
We assume $c=1$.
The function $H$ must satisfy only Eq. (\ref{64}). As already asserted, 
the dependence of $H$ on the retarded time $(-u)$ is arbitrary. The geodesic
equations in terms of the $t, x, y, z$ coordinates are

\begin{equation}
2\ddot{t} + \sqrt{2}H\ddot{u} + \sqrt{2}\dot{H}\dot{u} 
- {1\over \sqrt{2}}{\partial H\over \partial u}\dot{u}^2 = 0,
\label{66}
\end{equation}
\begin{equation}
2\ddot{x} - {\partial H\over \partial x}\dot{u}^{2} = 0,
\label{67}
\end{equation}
\begin{equation}
2\ddot{y} - {\partial H\over \partial y}\dot{u}^{2} =  0,
\label{68}
\end{equation}
\begin{equation}
2\ddot{z} +  \sqrt{2}H\ddot{u} + \sqrt{2}\dot{H}\dot{u} 
- {1\over \sqrt{2}}{\partial H\over \partial u}\dot{u}^2 = 0\,,
\label{69}
\end{equation}
where the dot represents derivative with respect to the parameter $s$ that
appears on the left hand
side of Eq. (\ref{65}). Among the various possibilities for the function
$H(-u)$, we have \cite{Maluf6}

\begin{equation}
H_1=A_{+}(u)(x^2-y^2)\,,
\label{70}
\end{equation}

\begin{equation}
H_2=A_{\times}(u)\,xy\,,
\label{71}
\end{equation}
or by a linear combination of these quantities. 
The two expressions above of $H$ satisfy Eq. (\ref{64}), and have been
considered in the literature in several investigations. We anticipate that
both expressions
lead to the same qualitative behaviour for the geodesics and kinetic energy 
of the particles. We consider the amplitudes 
$A_{+}(u)$ and $A_{\times}(u)$ to be given by regular Gaussians
(normalised, not necessarily to 1), that 
represent short bursts of gravitational waves. These amplitudes are typically
(but not exactly) given by

\begin{equation}
A_{(+,\times)} \simeq
{1\over \lambda }\; e^{-{(u^2/ 2\lambda^2})}\,.
\label{72}
\end{equation}
In addition to the ansatz above, we have chosen multiplicative constants in 
the expression of the amplitude in order to yield satisfactory pictures 
(see below). In fact, the amplitude does not need to be normalized.

The effect of a gravitational wave on a free particle is assumed to be very
weak. For this reason, the kinetic energy per unit mass $K$ of the free 
particles is written as

\begin{equation}
2K={1\over 2}(V_x^2 + V_y^2+ V_z^2)\;,
\label{73}
\end{equation}
where the velocities are
$V_x=dx/du$, $V_y=dy/du$ and $V_z=dz/du$. Thus, in view of Eq. (\ref{62}),
we have

\begin{equation}
K={1\over 2}\biggl[ \biggl({{dx}\over {dt}}\biggr)^2+ 
\biggl({{dy}\over {dt}}\biggr)^2+\biggl({{dz}\over {dt}}\biggr)^2 \biggr]\,,
\label{74}
\end{equation}
which is a valid expression for instants of time before and after
the passage of the wave, in which case the space-time is flat, or nearly 
flat.

We define $K_f$ and $K_i$ as the final and initial kinetic energies of the 
free particles, i.e., the energy of the particles after and before the
passage of the wave. We also define $\Delta K=K_f-K_i$, and the normalized
quantity $\Delta K_N$,

\begin{equation}
\Delta K_N= {{K_f-K_i}\over {K_f+K_i}}\,.
\label{75}
\end{equation}

The geodesic equations have been solved numerically, as explained in Ref.
\cite{Maluf6}, and the quantities $\Delta K$ and $\Delta K_N$ have also been
evaluated numerically, which resulted in the figures below. The initial 
conditions for the constructions of these figures were arbitrarily chosen, 
and are clearly presented in \cite{Maluf6}.

\begin{figure}
	\centering
		\includegraphics[width=0.80\textwidth]{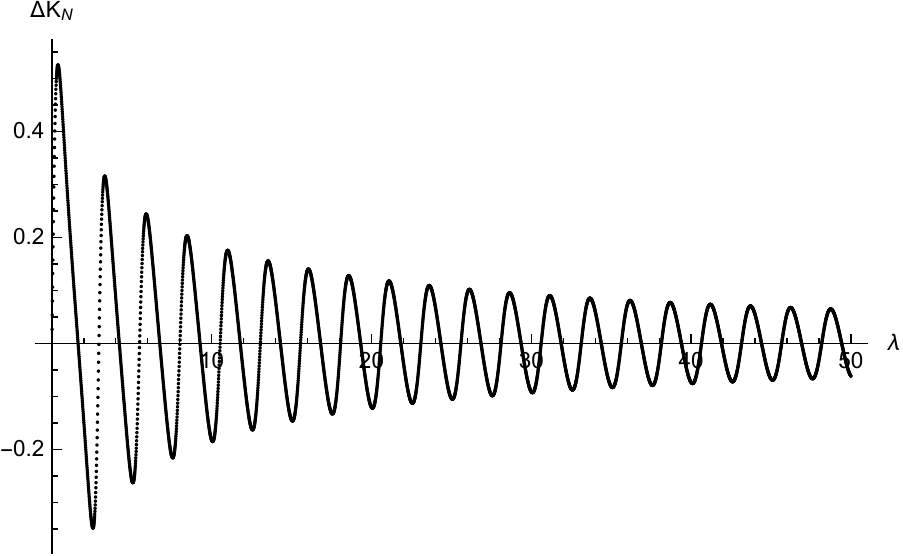}
	\caption{$\Delta K_N$ as a function of $\lambda$, considering $H_1$}
	\label{figure3}
\end{figure}

\begin{figure}
   \begin{minipage}{0.49\linewidth}
     \centering
     \includegraphics[width=\textwidth]{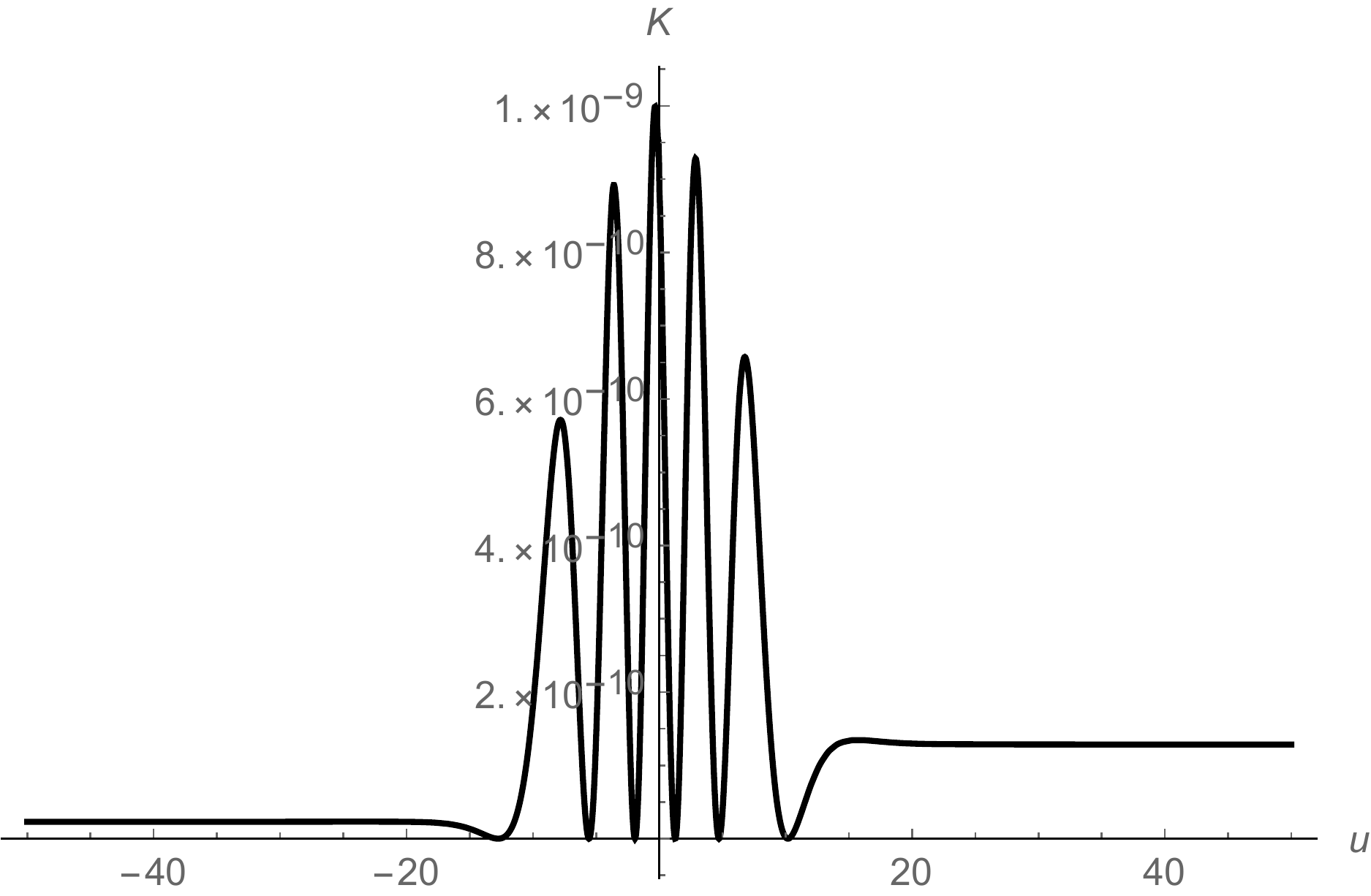}
     \caption{$K(u)$, considering $H_1$ and
          $\lambda =7.30247$.}\label{figure4}
   \end{minipage}\hfill
   \begin {minipage}{0.49\linewidth}
     \centering
     \includegraphics[width=\textwidth]{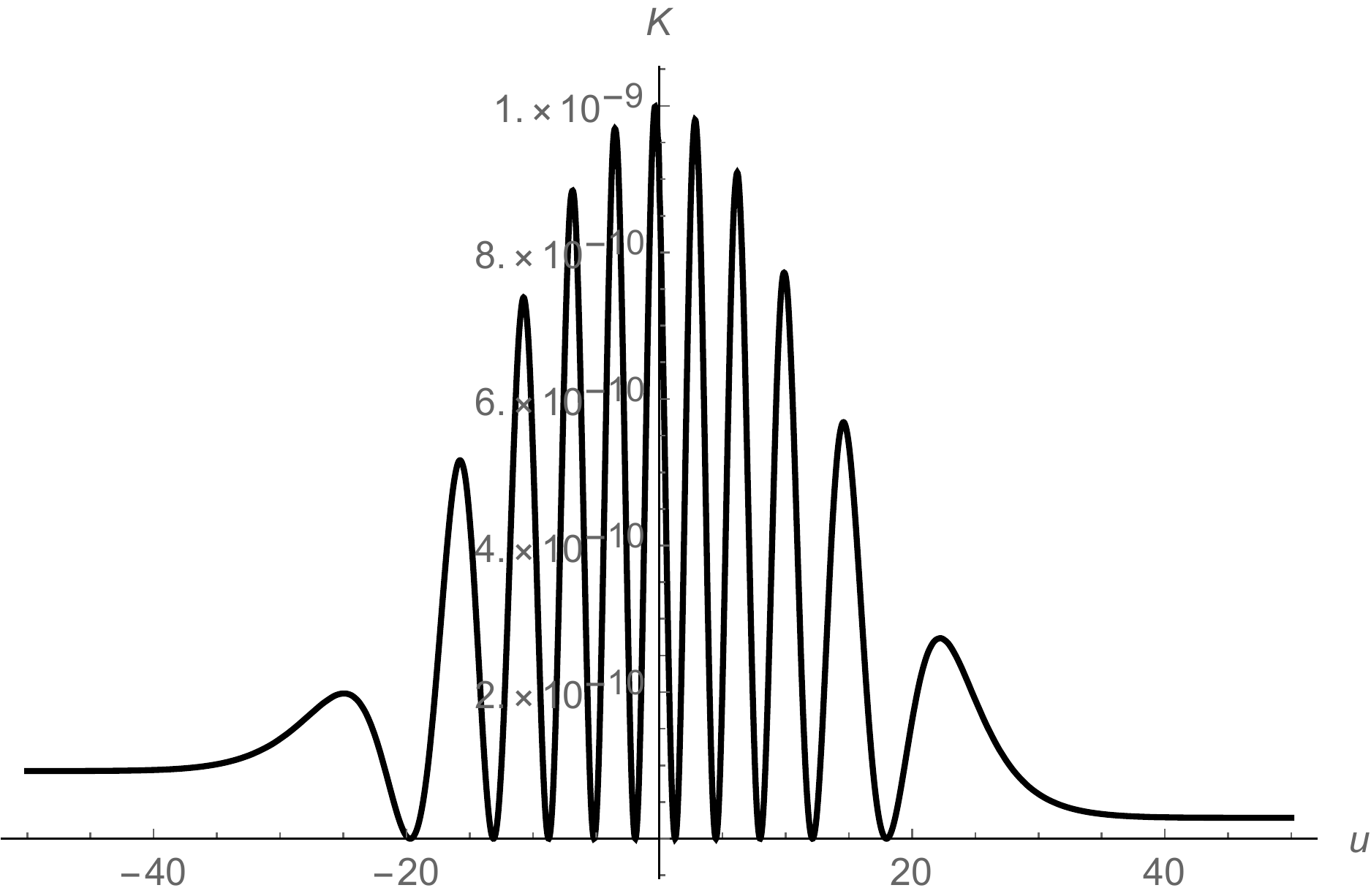}
     \caption{$K(u)$, considering $H_1$ and
          $\lambda =13.6043$.}\label{figure5}
   \end{minipage}
	\end{figure}
In Figure 3, we see an interesting variation of the normalised kinetic energy
with respect to the parameter $\lambda$ that appears in Eq. (\ref{72}). But
most important is the behaviour of $K(u)$ with respect to the variable 
$u$, which is given in Figures 4 and 5.

In Figure 4, the particle looses energy to the gravitational field, and the
opposite takes place in Figure 5. We have analysed innumerous situations 
where this energy transfer occurs, and that yield figures much different from
Figures 4 and 5. The gain or loss of energy depends
essentially in the initial conditions for the geodesic equations. However, 
an explanation to the variation of the kinetic energy of the free particle 
may be given by the 
work-energy relation for the free particle in the gravitational field. 

In Ref. \cite{Maluf7} we have investigated, in great detail, the validity of
the classical Torricelli equation, $v^2_f=v^2_i+2a\Delta x$, for a particle
that is subject to the gravitational field of a pp-wave. For a given 
arbitrary trajectory between the initial and final positions denoted by $i$ 
and $f$, respectively, the Torricelli equation is written as

\begin{equation}
{1\over 2} v^2_f-{1\over 2}v^2_i = \int^f_i{\bf a}\cdot d{\bf l}\,,
\label{76}
\end{equation}
where ${\bf a}$ is the acceleration of the particle due to the gravitational
field. The space-time at the initial and final positions of the particle is
assumed to be flat. The relation above is just a manifestation of the 
standard work-energy relation $\Delta K = \Delta W$ of classical Newtonian
physics, for a particle that is under the action of a force 
${\bf F}=m{\bf a}$. However, in Eq. (\ref{76}), both quantities $\Delta  K$
(the left hand side) and $\Delta W$ (the right hand side of (\ref{76})) are 
quantities per unit of mass. As we discussed in Ref. \cite{Maluf7}, the 
relativistic form of the right hand side of the equation above is calculated
according to the expression

\begin{equation}
\Delta W=-\int_i^f {\phi_{(0)(i)}e^{(i)}\,_{j} dx^{j}}=-\int_i^f 
a_j dx^j\,. 
\label{77}
\end{equation}

The procedure for calculating the right hand side of the equation above is 
the following. First, we establish a set of tetrad fields adapted to static 
observers in the space-time of a pp-wave given by 
the metric tensor (\ref{65}).
Then, we calculate the inertial accelerations that are
necessary to maintain the frame static in space-time. We identify the 
gravitational acceleration on the particle (or on any other physical body)
as exactly {\it minus} the 
inertial accelerations, since the frame must be static.
This identification explains the minus sign in Eq.
(\ref{77}), and in this equation $e^{(i)}\,_j$ are components of the static
co-frame. The acceleration tensor components $\phi_{(0)(i)}$ are calculated 
out of Eq. (\ref{1-5}) or (\ref{1-10}).

As for $\Delta K$, we used

\begin{eqnarray}
\Delta K &=&\frac{1}{4}\biggl[ \dot{x}^2(u)+\dot{y}^2(u)
+\dot{z}^2(u) \biggr]_{u=-\infty}^{u=\infty} \nonumber \\
&=&\frac{1}{2}\biggl[ \biggl( {{dx}\over{dt}}\biggr)^2 +
\biggl( {{dy}\over{dt}}\biggr)^2+
\biggl( {{dz}\over{dt}}\biggr)^2 \biggr]_{t=-\infty}^{t=\infty}\,.
\label{78}
\end{eqnarray}
The trajectories and initial conditions for the solution of the geodesic
equations are presented in detail in Ref. \cite{Maluf7}. 
We reaffirm that in the initial ($t\rightarrow -\infty$) and final
($t\rightarrow +\infty$) positions of the particle, the space-time is
flat, in view of the Gaussian form of the wave, and this fact means that
the initial and final geodesic trajectories of the particle are
straight lines in space-time.
By means of numerical and graphical analysis, we have found a striking
and astonishing coincidence between $\Delta K$ and $\Delta W$, for various
trajectories and initial conditions. 

A more realistic treatment of this gravitational field configuration would
take into account the contribution of the mass of the particle to the total
gravitational field. However, such a solution of the field equations does
not exist, even in approximate form. For a particle with very small mass,
that generates a gravitational field negligible compared to the field of the
wave, the geodesic equations considered in 
Refs. \cite{Maluf4,Maluf5,Maluf6,Maluf7} are consistent and realistic 
trajectories of the particles and lead to a measurable velocity memory 
effect. An interesting consequence of the configuration pp-wave + free 
particles is that it is possible, at least in principle, to extract energy
from gravitational waves, keeping in mind that the waves may as well carry 
away energy from material configurations.

The general conclusion is that we may have $\Delta K>0$
or $\Delta K <0$ depending {\bf (i)} on the initial and final conditions
of the geodesic trajectories, and {\bf (ii)}
on the gravitational acceleration on the particle,
due to the pp-wave. In any case, there is an actual transfer of energy
between the gravitational field of the pp-wave and the particle. This is
a possible explanation to the memory effect, in the context of non-linear
plane gravitational waves. 

One important conclusion of the present
review is that the energy transfer between the particle and the wave 
(i.e., the gravitational field) is pointwise, which demonstrates the 
localizability of the gravitational energy.

\section{Final Remarks}

In this review we have revisited the definitions of 
energy-momentum $P^a$ (Eq. (\ref{11})) and 4-angular momentum $L^{ab}$ 
(Eq. (\ref{15})) of the gravitational field in 
the TEGR, with emphasis on the localization of the gravitational energy
and, by extension, on the localization of the gravitational momentum and 
4-angular momentum. We argue that the velocity memory effect is a 
manifestation of the
localization of the gravitational energy: the transfer of energy between an
otherwise free particle and the gravitational field of a plane-fronted 
gravitational wave is pointwise, and so the gravitational energy must be
localized at the position of the particle. Therefore, the densities that 
arise in the TEGR and that yield the definitions of $P^a$ and $L^{ab}$ do
make physical sense. In addition, the results obtained in the analysis of
distinct configurations like the Kerr black hole, the Schwarzschild-de Sitter
space-time and the Bondi-Sachs space-time support the validity of $P^a$, 
given that all results are consistent from the physical point of view. These
space-times have specific features, and the definition of $P^a$ is in
agreement with the expected phenomenology regarding each one of these 
features, which is an indication of the universality of $P^a$. 

In Ref. \cite{Maluf18} we investigated the transfer of angular
momentum between a free particle and the gravitational field of a 
gyratonic pp-wave. Exactly as in the case of the energy transfer, the final
angular momentum of the particle may be smaller or higher than the initial
angular momentum. This variation of angular momentum of the particle depends
on the initial conditions of the particle, as well as on the nature of the 
gyratonic wave.

We have also revisited the concept of tetrad fields, and by means of the 
analysis of well known and physically motivated constructions, we have shown,
once again, that a set of tetrad fields in space-time yields the reference
frame of distinguished observers. The important and inevitable conclusion is
that tetrad fields describe at the same time the reference frame of a field
of observers, and the gravitational field. These two manifestations are 
inextricably linked. As we explained in Section 2, the set of tetrad fields
may be characterized {\bf (i)} either by fixing the timelike component of the
frame $e_{(0)}\,^\mu$ along the trajectory of an observer, and fixing the 
three other spacelike components according to the symmetry of the physical
configuration, or {\bf (ii)} by requiring the six components 
of the acceleration tensor to satisfy some conditions 
on the frame in space-time (as, 
for instance, to maintain the frame static and non-rotating in space-time).

Two sets of tetrad fields that are related by a local Lorentz transformation
may be equivalent in the sense that they yield the same metric tensor, but in
general they describe quite distinct 
congruences of observers (i.e., congruence of 
timelike curves defined by $e_{(0)}\,^\mu(x)$),
and yield different results for the gravitational
energy-momentum, which is covariant under global, but not under 
local Lorentz transformations. We have made clear in Sections 2 and 4 that 
the sets of tetrad fields given by Eqs. (\ref{1-5-4}) and (\ref{5-1-7}) 
for the Kerr line element are
physically different. Only the latter one has physical relevance for the 
result in Section 4, namely, to the evaluation of the irreducible mass of the
Kerr black hole.

The calculation of the energy contained within the
external event horizon of the Kerr black hole leads to a result that is
strikingly close to twice the irreducible mass of the black hole, which is 
the expected result. The final
expression obtained in our analysis is a lot more satisfactory than the 
result achieved by any other approach in the existing literature on this
subject. We must choose the set of tetrad fields adapted to the 
field of observers that are allowed within the ergosphere of the Kerr black 
hole. This set of tetrad fields is given by Eq. (\ref{5-1-7}). The frame 
described by the latter equation yields a field of observers that are dragged
in rotational motion by the gravitational field of the black hole inside the 
ergosphere, which is a well known feature of the Kerr space-time. On the 
other hand, tetrads defined by Eq. (\ref{1-5-4}) are not defined in the 
interior of the ergosphere, and for this reason
they play no role in the evaluation of the irreducible mass.

In order to present our definitions of $P^a$ and $L^{ ab}$, we have 
readdressed the Lagrangian and Hamiltonian approaches to the TEGR.
The Hamiltonian formulation 
is always important because it leads to the constraint equations and
guarantees the mathematical and physical consistency of the theory. 
The TEGR is a theory with only first class constraints, which
means that the hyperbolic evolution equations are well defined. The 
definitions of energy, momentum and 4-angular momentum of the gravitational
field in the TEGR are essentially a natural development of the Hamiltonian 
formulation. As we wrote before, they are not abstract mathematical 
elaborations motivated by circumventing a non-existent problem of 
localization of the gravitational energy.

The TEGR, as exposed in this review, is constructed out of the tetrad fields
only. The addition of a (flat or non-flat) spin connection does not change 
the dynamics of the theory, and 
just brings unnecessary complications both for the mathematical evolution of
the Hamiltonian field equations and for the measurable
quantities of the theory. There is, currently, a debate in
the physics community regarding this issue, as we can see in the exposition
of Refs. \cite{Hohmann1,Hohmann2,Blixt,Golovnev}, and references therein. 
We also
call attention to the interesting considerations carried out in Ref. 
\cite{Formiga6}, regarding the establishment of tetrad fields in space-time.
The notion of ideal or preferred frames has always been pursued in the
context of tetrad theories of gravity, of which the TEGR is one prominent 
realisation.

The Lagrangian and Hamiltonian formulations of the TEGR considered in this
review are invariant under the global SO(3,1) group. However, the field 
equations of the theory, Eqs. (\ref{5-1}) or (\ref{5}), as discussed 
before, are covariant under
the local SO(3,1) group. The fact that the field equations are covariant 
under local Lorentz transformations  means that the theory,
defined by the field equations, can
be formulated in the frame of an arbitrary observer in space-time. This is
the essential and crucial meaning of local Lorentz invariance of a 
theory. Measurable 
quantities, of course, depend on the frame. Even in the standard formulation
of classical electrodynamics, the electric field of a point charge has an
expression in a static frame (where the charged particle is at rest),
and a different expression in a frame that is moving with respect to the 
point charge. Assuming that the space-time is flat and that the velocity of
the moving frame is constant, both observers at the static and moving frames
move along geodesics in space-time. In spite of moving along geodesics, they
measure different values for the electric field, and this is just a 
consequence of the special theory of relativity. It simply does not make 
sense to allow this feature to take place in classical electrodynamics, and 
not in context of a theory for the gravitational field. In the framework of
the special theory of relativity, physical measurements are frame dependent,
and so they should be in the realm of general relativity. After all, one of
the attributes of tetrad fields is exactly to project vectors and tensor
components (coordinate dependent quantities) into the local frame of an 
observer.

\end{document}